\documentclass[modern]{aastex61}

\usepackage{color}

\received{April 4, 2018}
\revised{April 30, 2018}
\accepted{April 30, 2018}

\submitjournal{ApJ}

\shorttitle{Limits of K2 M67 Rotation Period Measurements}
\shortauthors{Esselstein et al.}

\begin{document}

\bibliographystyle{aasjournal}

\title{The K2 M67 Study:  Establishing the Limits of Stellar Rotation Period Measurements in M67 with K2 Campaign 5 Data}

\correspondingauthor{Rebecca Esselstein}
\email{rebecca.esselstein@physics.ox.ac.uk}

\author[0000-0002-0786-7307]{Rebecca Esselstein}
\affiliation{Department of Astrophysics, University of Oxford \\
Denys Wilkinson Building, Keble Rd. \\
Oxford, OX1 3RH, UK}

\author{Suzanne Aigrain}
\affiliation{Department of Astrophysics, University of Oxford \\
Denys Wilkinson Building, Keble Rd. \\
Oxford, OX1 3RH, UK}

\author{Andrew Vanderburg}
\affiliation{NASA Sagan Fellow}
\affiliation{Department of Astronomy, University of Texas at Austin \\ 2515 Speedway, Stop C1400 \\ Austin, TX 78712-1205, USA}
\affiliation{Harvard-Smithsonian Center for Astrophysics \\ 60 Garden St. \\ Cambridge, MA 02138, USA}
%
\author{Jeffrey C. Smith}
\affiliation{SETI Institute \\
189 Bernardo Ave. Suite 100 \\
Mountain View, CA 94043, USA}
\affiliation{NASA Ames Research Center \\
Moffett Field, CA 94035, USA}
%
\author{Soren Meibom}
\affiliation{Harvard-Smithsonian Center for Astrophysics \\ 60 Garden St. \\ Cambridge, MA 02138, USA}
%
\author{Jennifer Van Saders}
\affiliation{Institute for Astronomy - M\=anoa, University of Hawai`i \\ 2680 Woodlawn Dr. \\ Honolulu, HI 96822, USA}
%
\author{Robert Mathieu}
\affiliation{Department of Astronomy, University of Wisconsin-Madison \\ 475 N Charter St. \\ Madison, WI 53706, USA}

\begin{abstract}
The open cluster M67 offers the unique opportunity to measure rotation periods for solar-age stars across a range of masses, potentially filling a critical gap in the understanding of angular momentum loss in older main sequence stars.  The observation of M67 by NASA K2 Campaign 5 provided light curves with high enough precision to make this task possible, albeit challenging, as the pointing instability, 75\,d observation window, crowded field, and typically low-amplitude signals mean determining accurate rotation periods on the order of 25 -- 30\,d is inherently difficult.  Lingering, non-astrophysical signals with power at $\geq$25\,d found in a set of Campaign 5 A and F stars compounds the problem.  To achieve a quantitative understanding of the best-case scenario limits for reliable period detection imposed by these inconveniences, we embarked on a comprehensive set of injection tests, injecting 120\,000 sinusoidal signals with periods ranging from 5 to 35\,d and amplitudes from 0.05\% to 3.0\% into real Campaign 5 M67 light curves processed using two different pipelines.  We attempted to recover the signals using a normalized version of the Lomb-Scargle periodogram and setting a detection threshold.  We find that while the reliability of detected periods is high, the completeness (sensitivity) drops rapidly with increasing period and decreasing amplitude, maxing at 15\% recovery rate for the solar case (i.e. 25\,d period, 0.1\% amplitude).  This study highlights the need for caution in determining M67 rotation periods from Campaign 5 data, but this can be extended to other clusters observed by K2 and, soon, TESS.
\end{abstract}


\section{Introduction} \label{sec:intro}

The measurement of rotation periods in open clusters can be a powerful and convenient tool in understanding stellar angular momentum evolution by offering the opportunity to connect a star's age to its rotation period and color or, alternatively, mass.  \citet{1972ApJ...171..565S} first noticed the relationship between period and age, though the color-period relation has long been observed in the Hyades, where there is a notable dependency of stellar rotation period on spectral type for stars cooler than about mid-F \citep{1987ApJ...321..459R,2011MNRAS.413.2218D}.  \citet{1988ApJ...333..236K} concluded that by the age of the Hyades, a star's initial angular momentum no longer plays a significant role in determining its rotation period, and another mechanism, possibly stellar winds coupled with magnetic fields, `brakes' the rotation of a star as it ages.  \citet{1989ApJ...343L..65K} then acknowledged the use of rotation, a direct observable independent of distance, as a potential age estimator. \citet{2003ApJ...586..464B} developed the first semi-empirical model for deriving ages from the colors and rotation periods of FGK dwarfs, introducing the term gyrochronology. Following the improvements of \citet{2007ApJ...669.1167B}, there was a need for precise rotation period measurements in clusters to test gyrochronology and probe its limitations.  The availability of large field-of-view (FOV) CCD cameras on smaller (1\,m class) telescopes enabled the measurement of thousands of rotation periods for stars in young open clusters via the rotational modulation of stellar flux \citep{irwin2007monitor,irwin2009monitor,2009ApJ...695..679M,0004-637X-733-2-115,2009MNRAS.400..451C,0004-637X-695-1-336,2010arXiv1006.0950H,2010A&A...515A.100J}, using co-eval populations to enable the creation of period-age-mass (P-t-M) surfaces based on different gyrochronology models.  

NASA's \emph{Kepler} mission, dedicated to the detection of transiting exoplanets of nearby bright stars, offered unprecedented precision to study stellar variability, particularly stellar rotation.  Key \emph{Kepler} large-scale rotation period studies include those of \citet{2013MNRAS.432.1203M}, \citet{2013A&A...560A...4R},  \citet{2041-8205-775-1-L11} (letter), \citet{2014ApJS..211...24M}, \citet{2013A&A...557L..10N}, \citet{2014A&A...572A..34G}, and \citet{2015A&A...583A..65R}. These studies measured the rotation periods of tens of thousands of \emph{Kepler} field stars, including planetary hosts. Along with the ground-based cluster surveys, they helped bring to light a number of interesting trends, including, for example, the puzzling bimodal period distribution shown by late K and M-dwarfs, as well as the existence of a correlation between stellar temperature, rotation period, and the spread thereof, which may be related to differential rotation, active region evolution, or a combination of the two.

The Kepler Cluster Study extended the P-t-M surface from $\sim$600\,Myr (Hyades) to 2.5\,Gyr by using \emph{Kepler} data to measure rotation periods of FGK dwarfs in NGC 6811 (1\,Gy;\citealt{2011ApJ...733L...9M}) and NGC 6819 (2.5\,Gyr;\citealt{2015Natur.517..589M}).  These measurements agreed with the predictions of the most recent gyrochronology relations by \citet{0004-637X-722-1-222}.  Other approaches using \emph{Kepler} data, however, have focused on asteroseismic age determination \citep{2014A&A...572A..34G,2015MNRAS.450.1787A,2016Natur.529..181V}, and these studies have called into question the behavior of stars at older ages.  For instance, \citet{2016Natur.529..181V} attempts to explain the anomalously rapid rotation of old field stars by proposing a `weakened magnetic braking' theory, which states that around solar age, stars with masses similar to and above that of the Sun should begin to diverge from the `standard' gyrochronology relations, which their cooler counterparts are still expected to follow.  The need for high precision observations of solar age clusters to test this new theory alongside the established gyrochronology relations becomes apparent.

The best candidate for an in-depth rotation study on a solar-age cluster is M67.  M67 is estimated to be $\sim4.3$\,Gyr old, and roughly 800 -- 900\, pc from the Sun \citep{1994A&AS..103..375C,2007A&A...470..585B,2015AJ....150...97G}.  The cluster's age is not the only similarity to the Sun; calculated metallicty values from the literature show that [Fe/H] ranges from about $-0.1$ to $0.1$, with average values lying around $-0.01$ \citep[][the most recent of which reports $-0.01 \pm 0.05$]{1996AJ....112..628F,2007A&A...470..585B,2007AJ....133..370T,2011AJ....142...59J}. The members of M67 have been catalogued several times based on proper motion \citep{1977A&AS...27...89S,2005ARep...49..693L,2008A&A...484..609Y} and radial velocities \citep{2008A&A...484..609Y,2008A&A...489..677P,2015AJ....150...97G}. \citet[][hereafter G15]{2015AJ....150...97G} report 1278 candidate members, making M67 the richest, relatively nearby solar age open cluster.  
There is a general tendency for middle-aged, main-sequence stars to have longer activity cycles and so lower variability amplitudes than their younger counterparts \citep{2008ApJ...687.1264M, 1995ApJ...452..332R, 1998ApJS..118..239R, 1996A&A...305..284H, 1998ASPC..154..153B}, making their signals more difficult to detect.  Thus high photometric precision is necessary for a study of rotation in M67.  K2, the second phase of the \emph{Kepler} mission following the loss of two of the spacecraft's reaction wheels by 2013, presents such and opportunity.  However, the mission's reduced pointing accuracy relative to \emph{Kepler} leads to lower photometric precision due to an increase in systematics associated with inter- and intra-pixel sensitivity variations, as well as aperture losses \citep{2014PASP..126..398H}.  Early estimates of K2's photometric precision indicated, for stars with a magnitude of $V=12$, a precision of approximately 400 parts per million (ppm) for the long-cadence, or 30-minute, observations, and 80 ppm over the course of 6 hours \citep{2014PASP..126..398H}.  Previously, \emph{Kepler} reached a precision level of 10 ppm at a magnitude of $K_{p}=10$ for 6-hour observations \citep{2012PASP..124.1279C, 2014PASP..126..948V}. 

Members of the community have since developed methods to improve the photometric precision of the K2 light curves, in particular to correct the pointing-related variations. These include the Vanderburg \& Johnson (hereafter VJ) pipeline \citep{2014PASP..126..948V,2016ApJS..222...14V}, the K2 Systematics Correction (K2SC) pipeline \citep{2015MNRAS.447.2880A,2016MNRAS.459.2408A}, EVEREST \citep{2016AJ....152..100L}, K2VARCAT \citep{2014arXiv1411.6830A,2016MNRAS.456.2260A}, and the PSF-fitting used by \citet{2016MNRAS.456.1137L} for K2 observations of M35 and NGC 2158.  Several of these pipelines match the photometric precision of \emph{Kepler} for stars with $K_{p}=12.5$, and perform within a factor of two of the original mission for fainter stars. This, combined with the much more diverse sample of stars surveyed by K2 compared to the \emph{Kepler} prime mission, including numerous open clusters, makes K2 data a treasure trove for stellar angular momentum evolution studies. Indeed, the K2 mission has produced spectacular results for nearby, well-studied young open clusters, including the Pleiades \citep{2016AJ....152..113R,2016AJ....152..114R,2016AJ....152..115S} and Hyades \citep{2016ApJ...822...47D}.  More critical to this study, however, K2 observed M67 during Campaign 5.\footnote{https://keplerscience.arc.nasa.gov/k2-fields.html} \citet{2016ApJ...823...16B} and \citet{2016MNRAS.459.1060G,2016MNRAS.463.3513G} already used those data to measure rotation periods for M67 members and investigate the implications of their results in terms of gyrochronology.  However, both studies worked only with stars observed using individual postage stamps, avoiding the crowded central regions of the cluster, and while \citet{2016ApJ...823...16B} obtained a good match to pre-existing gyrochronology relations, this was based on a very small sample of 20 stars, with the majority of the reported periods not convincingly confirmed when examined by eye based on the provided light curves.  By contrast, \citet{2016MNRAS.463.3513G} showed no dependence of rotation period on mass (color) and significant scatter for all masses.  Furthermore, there are discrepancies in the measured periods for several stars in common between the two studies, and they inferred mutually inconsistent gyrochronological ages for M67 ($4.2 \pm 0.2$ and $5.0 \pm 0.2$\,Gyr, respectively). 

\begin{figure*}
\includegraphics[width=1\linewidth]{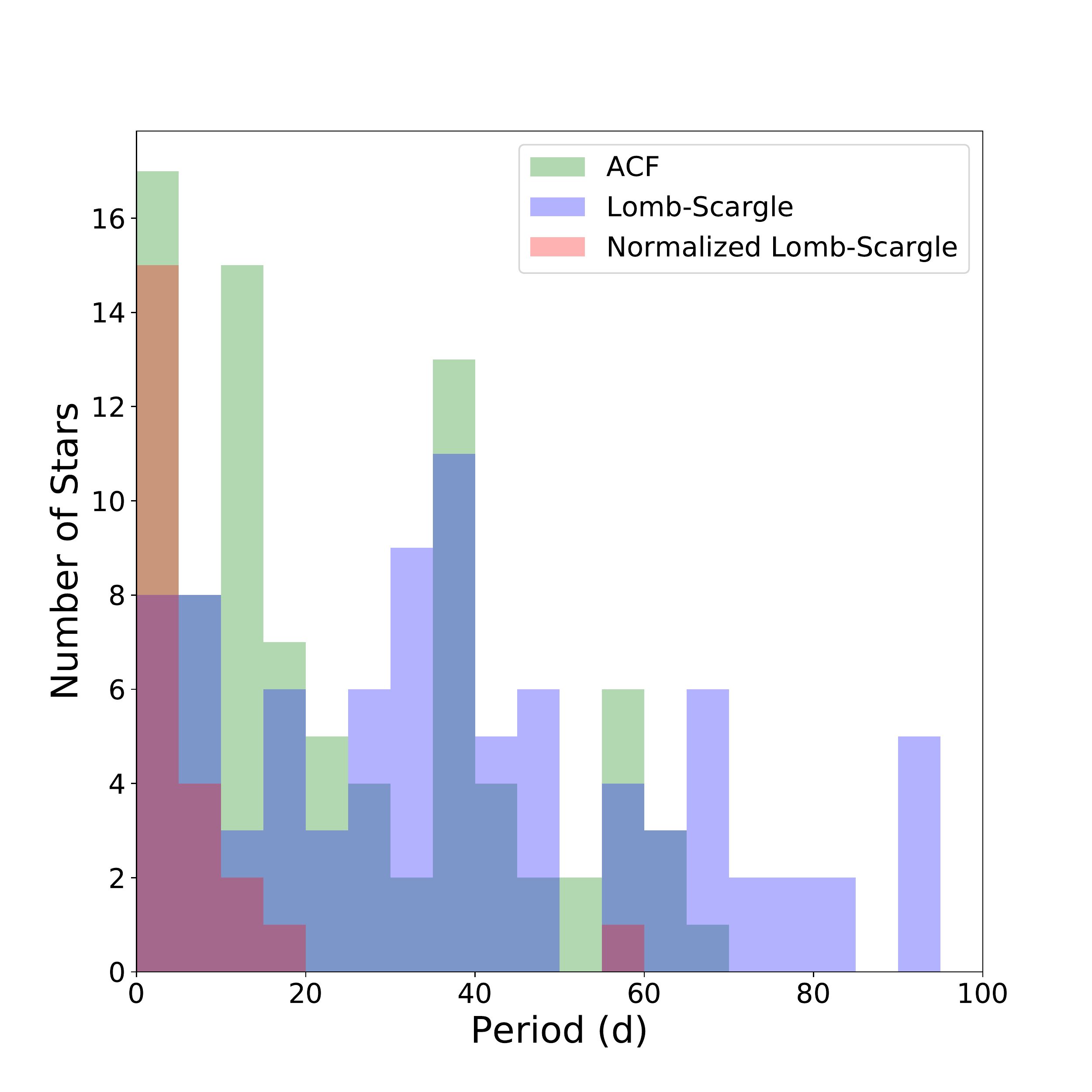} \caption{Periods acquired from a set of 89 K2 Campaign 5 A and F stars using the autocorrelation function (ACF) (green) and the Lomb-Scargle periodogram (blue).  We also include a normalized version of the Lomb-Scargle periodogram (red), which we implement later in this paper, as a demonstration of how we can set detection thresholds and remove potentially anomalous measurements. \label{fig:hotstarhist}}
\end{figure*}

In addition to exploiting only part of the data available, the published K2 M67 rotation studies made specific choices in terms of light curve extraction, detrending and period search methods, without investigating possible alternatives in any detail, and both methods involved a large component of human decision-making, which is difficult to reproduce. Furthermore, it is useful to dwell on just how challenging it is to determine accurate rotation periods in M67 data. Based on existing gyrochronology relations, and on what we know about the Sun itself, these stars are expected to have rotation periods of order 20 -- 30\,d, and amplitudes of a few tenths of a percent at most. Even in the full \emph{Kepler} field star sample of \citet{2014ApJS..211...24M} based on 4 years of continuous observations, there are relatively few detections in this area of parameter space, due either to sensitivity limits or the underlying period-amplitude distribution.  The limited duration of a K2 campaign ($\sim$75\,d; about 2 -- 3 rotation cycles for a typical M67 star) will only make the process of detecting accurate rotational modulation harder.  In addition, the reduced photometric precision of K2 light curves (relative to the \emph{Kepler} prime mission) is compounded by the significant degree of crowding of cluster stars, which creates additional systematics, and the large \emph{Kepler} pixel size.

In addition to the difficulties mentioned above, a sample of 89 hot A and F stars from K2 Campaign 5 has highlighted the presence of 25\,d$+$, non-astrophysical signals in the data, as can be seen in Figure~\ref{fig:hotstarhist}.  This figure shows the rotation periods of the dataset, acquired using the autocorrelation function (ACF), the standard Lomb-Scargle periodogram, and a normalized version of the Lomb-Scargle periodogram (which we explain and implement in this study in Section~\ref{sec:tests}).  The hot star sample selection is detailed in Section~\ref{subsec:star_samples}.  We should not expect periods much greater than $\sim$5\,d for these stars.  The significant number of measurements from either the ACF or the standard Lomb-Scargle periodogram above 25\,d shows that there is a substantial risk of making erroneous period measurements on the order of 25 -- 30\,d in M67.  All of this should lead to a healthy degree of prior skepticism regarding any period detections in M67, for G and K stars at least, and any such detections need to be backed up by detailed completeness (sensitivity) and reliability tests.  

Using all the cluster members and the aforementioned set of A and F stars observed during K2 Campaign 5, we carefully compare the effects of different light curve extraction and detrending methods. We perform systematic tests with simulated, sinusoidal signals injected into the Campaign 5 M67 K2 light curves, in order to properly evaluate the biases and best-case sensitivity limits of each detrending pipeline.  This paper reports on the first set of injection tests carried out as part of our team's K2 M67 rotation study, which ultimately seeks to produce a list of periods from this critical cluster in which we can quantify our level of confidence.  Later, a round of smaller injection tests will use more realistic signals, generated by the star spot models from \citet{2015MNRAS.450.3211A}, to help us reach that final goal. 

The structure of the rest of this this paper is as follows:  section \ref{sec:lcs} discusses the K2 observations, light curve extraction, and light curve detrending.  Section \ref{sec:tests} describes the injection tests, including the generation of the injected signals, the period search method, and the definition of a systematic detection criterion. In this section we also discuss how we handle intrinsically variable stars (i.e. stars with an astrophysical signal before injection), and what we consider a `valid' detection. The results are presented in Section \ref{sec:results}, where we compute the completeness and reliability of the period search as a function of period and amplitude, before they are discussed in Section \ref{sec:discussion}. We summarize our conclusions in section \ref{sec:conclusion}. 

\section{Observations and light curve preparation} 
\label{sec:lcs}

\subsection{K2 Observations}

\begin{figure}
\plotone{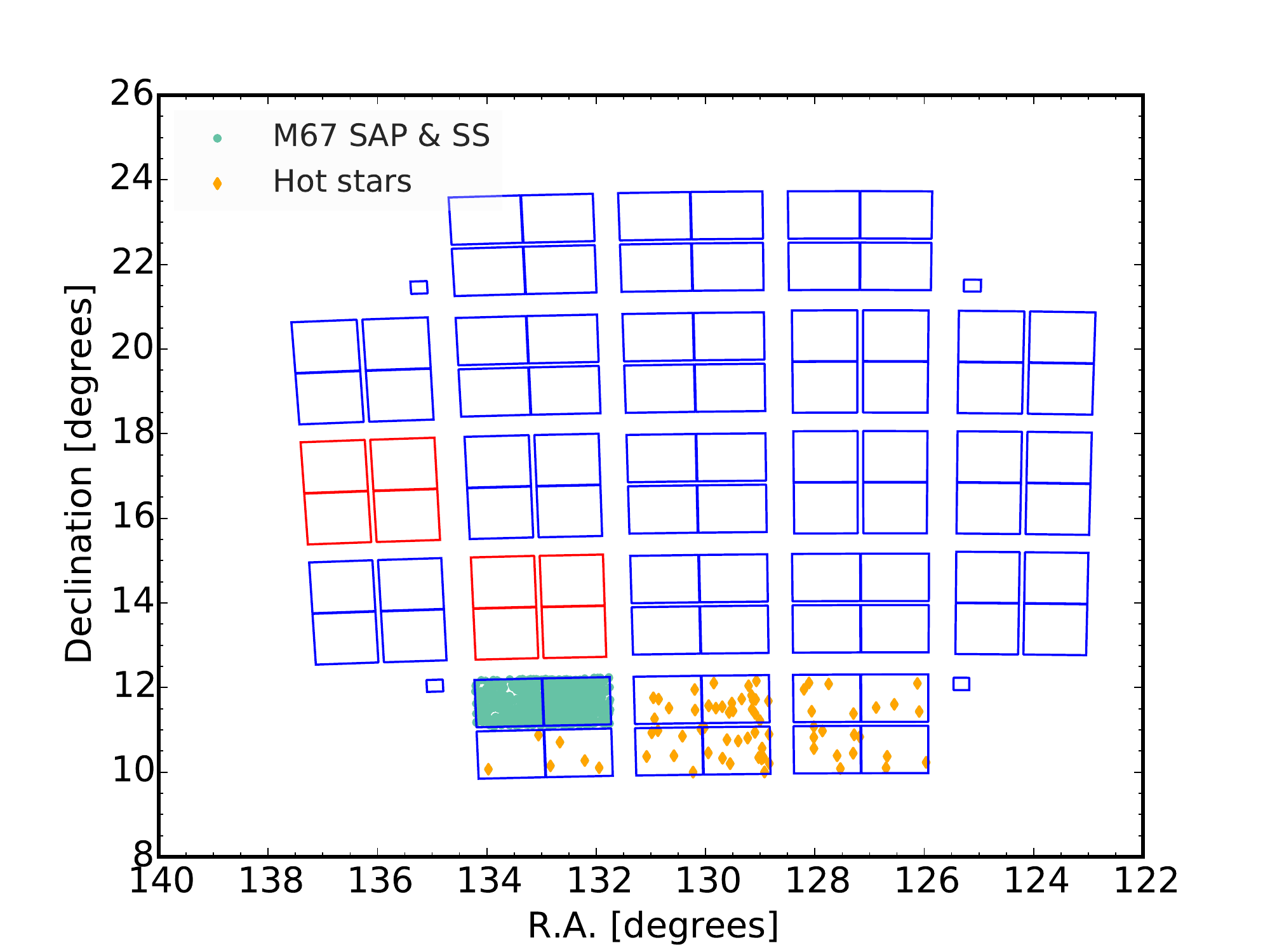}
\caption{Location of the 1877 M67 SAP and 976 superstamp stars (teal circles) used in this study, along with an additional 89 hot stars (orange diamonds), on the \emph{Kepler} Campaign 5 field of view.  \label{fig:mod6_fov}}
\end{figure}

M67 was observed during K2 Campaign 5, which commenced on 2015 April 27 and ended on July 10. The pointing was centred on RA = 130$^h$09$^m$27$^s$.53 and DEC = 16$^\circ$49'46.61''.\footnote{See \url{https://keplerscience.arc.nasa.gov/k2-data-release-notes.html\#k2-campaign-5}.} Individual K2 targets for Campaign 5 were proposed for observation by members of the community. Our proposal (PI.\ R.\ Matthieu) included $12\,935$ candidate members, with $12\,521$ drawn from the EPIC database, 266 from the 2MASS calibration field \citep[see e.g.][]{2009ApJ...698.1872S}, and the last 148 targets filling in those missing from the EPIC database around $K_{p}\sim19$ (Gilliland, private communication).  Membership for stars brighter than $K_{p}=15$ is based on radial velocity and proper motion, as explained in G15, taking those with membership probabilities of 20\% or greater to make a more complete catalog.  For the remainder, we accepted those stars whose photometric membership made them likely members, unless contradicted by known radial velocities or proper motions.  These stars formed our master target list.  M67 was located on channels 1 and 2 of CCD module 6, near the edge of the FOV. Figure~\ref{fig:mod6_fov} shows the entire Campaign 5 FOV, with M67 targets represented by the teal circles on CCD module 6 along with a scattering of orange diamonds that represent the sample of hot stars used in this analysis, which will be discussed later.

\begin{figure}
\centering
\includegraphics[width=1\linewidth]{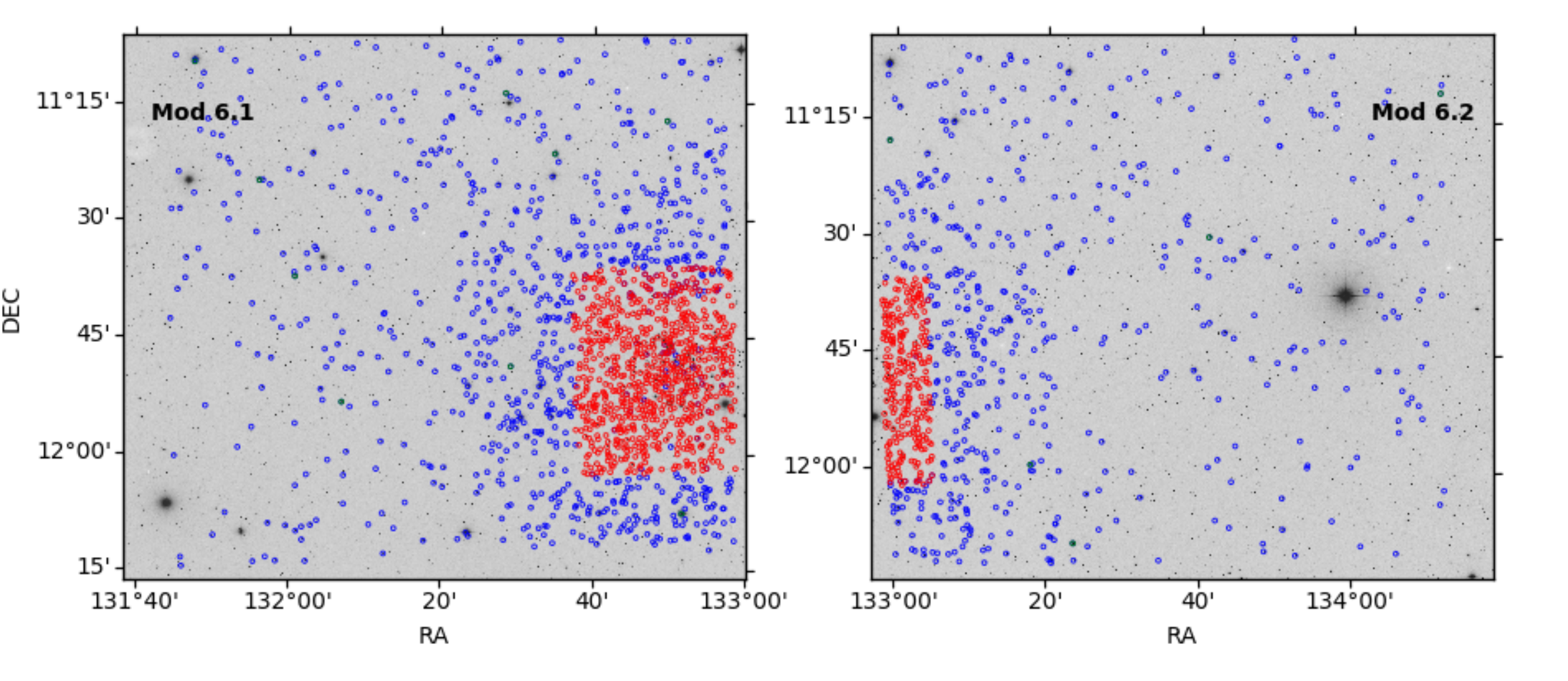}
\caption{Zoomed in look at the M67 SAP and superstamp stars using two images of M67 retrieved from the ESO Online Digitized Sky Survey.  The SAP stars are surrounded by blue circles while the SS stars are surrounded by red circles. \label{fig:mod6_fov2}}
\end{figure}

The stars located in the less crowded, outer parts of the cluster, were observed in the standard manner, by downloading a small window of pixels centred on each star of interest. A total of $\sim25\,000$ individual star postage stamps were collected during Campaign 5; of these 1877 fell on output channels 6.1 and 6.2.  For the crowded inner regions of the cluster, a `superstamp' was used: a large, contiguous region created by juxtaposing many postage stamps. Approximately 2210 of the proposed M67 targets were located within the superstamp (the number of stars in the superstamp for which light curves are actually extracted depends on the method used).  Figure~\ref{fig:mod6_fov2} shows the locations of the M67 SAP sample (in blue) and the superstamp stars (in red) where they fall on channels 6.1 and 6.2 using two images of M67 taken from the ESO Online Digitized Sky Survey, centered on the channels' respective pointing.

Most postage stamps were read out in `long cadence' mode, or one observation every 30 min, with a total of 3663 cadences during the campaign. After basic reduction (correction of pixel-level instrumental effects), each series of postage stamps was combined into 
a Target Pixel File (TPF), and the TPFs were made publicly available at the Mikulski Archive for Space Telescopes (MAST) a few months after the end of the campaign.

In the rest of this section, we describe the two parallel light curve  preparation procedures, or `pipelines', from two different institutions (Oxford and Harvard-CfA in collaboration with NASA Ames), which we tested against each other in this study.  The K2 M67 study team includes core members of the teams behind the K2SC and VJ pipelines, which we implement here.  While other K2 pipelines have been successful in various contexts, we did not include them in this study for a variety of reasons.  The EVEREST pipeline, for example, does not perform optimally with a crowded field \citep{2016AJ....152..100L,2017arXiv170205488L}.  Furthermore, the K2 light curves from K2VARCAT \citep{2014arXiv1411.6830A, 2016MNRAS.456.2260A} only encompass Campaigns 0 -- 4.  Alternatively, \citet{2018arXiv180206354C} have recently released Campaign 5 M67 light curves using a similar approach to superstamp data as \citet{2015MNRAS.447.2880A} (which we employ here).  \citet{2016MNRAS.463.1831N} have also produced Campaign 5 M67 light curves using the PSF-fitting approach of \citet{2016MNRAS.456.1137L}  in addition to aperture and optimal-mask photometry.  However, the light curves from \citet{2018arXiv180206354C} were unavailable when we started this study, and we did not realize those of \citet{2016MNRAS.463.1831N} were publicly available until after we completed it.  A future comparison with both sets of light curves would be interesting, especially since the root mean square error of the latter using aperture photometry at the bright end is similar to that of the pipelines we use (see Section 2.4).

Therefore, for each star, we applied our two pipelines to extract the light curves, correct for pointing-related systematics, and correct for residual common-mode systematics.  In an ideal world, one would try every possible combination of each of these steps, resulting in 8 versions of each light curve from the three separate datasets, each of which would have been injected with 42 simulated signals.  This would have led to an unmanageably large set of tests (almost 1\,000\,000 in total), and would have taken far too long to complete, especially when applying PDC-MAP for common-mode correction (see Section~\ref{subsubsec:cfa_commonmode}).  While mixing the pipelines may have produced slightly more optimal results, we treat each pipeline as a unit, and produced only two versions of each light curve.

\subsection{The `Oxford' pipeline}
\label{subsec:oxford}

The `Oxford' pipeline is comprised of the following:
\begin{itemize}
\item for light curve extraction: the Simple Aperture Photometry (SAP) component of the standard \textit{Kepler} pipeline, or the procedure described in \citet{2015MNRAS.447.2880A} for cases where the SAP light curves are not available;
\item for pointing systematics correction: the K2SC pipeline \citet{2016MNRAS.459.2408A};
\item for the correction of residual common-mode systematics: a Principal Component Analysis (PCA) step.
\end{itemize}

\subsubsection{Light curve extraction}
\label{subsubsec:extract1}

Stars located outside the superstamp were processed on the ground by the simple aperture photometry (SAP) component of the \emph{Kepler} 
pipeline. SAP light curves are extracted using a fixed, pixelated aperture, the boundaries of which are defined so as to maximize the signal-to-noise ratio of the resulting light curve \citep{2016PASP..128g5002V}. They are corrected for known, pixel-level instrumental effects only, and  contain significant instrumental systematics, primarily due to the pointing variations of the telescope. These light curves are publicly available at MAST. The MAST light curve files also include a second version of each light curve, produced by the Pre-search Data Conditioning (PDC) component of the \emph{Kepler} 
pipeline. The PDC step was designed to remove common-mode systematics and prepare the light curves for planetary transit searches. It was also designed to preserve all astrophysical signals, but it can reduce the amplitude of these intrinsic signals at timescales greater than 15\,d \citep{2015AJ....150..133G}. Additionally, in the case of K2 data, it was never designed to remove the sawtooth systematics associated with the spacecraft roll error \citep{1538-3873-124-919-985,1538-3873-124-919-1000,2014PASP..126..100S}.  Therefore, the use of PDC light curves can be problematic for variability studies with K2, so we work with the SAP rather than the PDC versions. 

As previously mentioned, MAST light curves are not available for the superstamp.  We therefore use the method of \citet{2015MNRAS.447.2880A} to extract the superstamp light curves ourselves.  We give a brief summary of it here.  We processed channels 6.1 and 6.2 in turn channel in turn.  First, a Full Frame Image (FFI) is downloaded and a preliminary astrometric solution obtained using {\tt astrometry.net} \citep{lang2010}. For each epoch, we then stitch together all the superstamp postage stamps from a given channel to create a reconstructed image, which is blank where no data was downloaded, and is initialized with the preliminary astrometric solution from the FFI. We also construct a binary mask that indicates where data was collected. We then use the publicly available {\tt CASUTools} routines {\tt nebuliser}, {\tt imcore}, and {\tt wcsfit} to model and subtract the background, identify stars within each image, and obtain a refined astrometric solution by cross-matching the catalog generated from each image with 2MASS. The variable background subtraction is new (compared to \citealt{2015MNRAS.447.2880A}) and is important because of the high density of unresolved stars in the central regions of the cluster. 

Next, we construct a master image by stacking the first 20 valid images using the astrometric solution for each frame. We repeat the source detection, photometry, and astrometric solution steps on this master image to obtain a master catalog. We then extract light curves for every star on the master catalog by placing a circular aperture at the sky position of each star on each image (as recorded in the master catalog).  We compute the median of each star's light curve and a zero-point correction for each frame by taking the weighted average of the median-subtracted fluxes, using the inverse square scatter as weights, on that frame. This removes some systematics (particularly those associated with aperture losses), but significant residual systematics are nonetheless present in the extracted light curves. 

The light curve extraction is done using several aperture radii (1, 2, 2$\sqrt{2}$, 4, 4$\sqrt{2}$, and 8 pixels) for every star in the superstamp.  In general, larger apertures give the best results for bright stars, while fainter stars require smaller apertures to minimize the background noise. However, the best aperture for any given star also depends on the relative positions and brightness of neighbouring stars. In practice, we select the aperture which minimizes the star's calculated scatter (see section \ref{subsec:p2p}).  Eventually, however, this aperture needs to be checked for contamination from other stars, especially if it is larger than about 2 pixels. 

We extracted light curves for 1342 stars located in the superstamp.  We then cross-matched our master catalog to that of G15, finding a total of 976 matches.  Most of the stars in our catalog without G15 matches are  faint stars with poor quality K2 photometry, so we focus the remainder of our analysis of superstamp stars on those with G15 matches, hereafter referred to as the `SS' sample. 

\subsubsection{Star-by-star systematics correction} 

The dominant systematics in K2 light curves are due to the spacecraft's pointing variations.  We correct these using the K2SC pipeline.  We briefly summarize K2SC here, but the more interested reader is referred to \citet{2016MNRAS.459.2408A}.  K2SC models each light curve as the sum of two unknown functions, one which depends on the star's pixel position and represents the pointing systematics, and one which depends on time and represents the star's intrinsic variations, as well as any residual systematics not captured by the position-dependent component. The model is formulated as an additive Gaussian process, which allows us to require only that each function have a certain degree of smoothness, without having to specify its exact form. It also enables us to separate the time- and position-dependent components.  Removing the former from the original light curve is helpful for looking for planetary transits, while subtracting the latter, which is dominated by the $\sim$6-hour drift of the spacecraft, only leaves us with a light curve corrected for position-dependent systematics, with minimal impact on astrophysical variability. K2SC also flags significant outliers in the data, which we subsequently remove since the rotational signals in which we are interested are relatively smooth for this study. 

\subsubsection{Common-mode systematics correction}

\begin{figure*}
\plotone{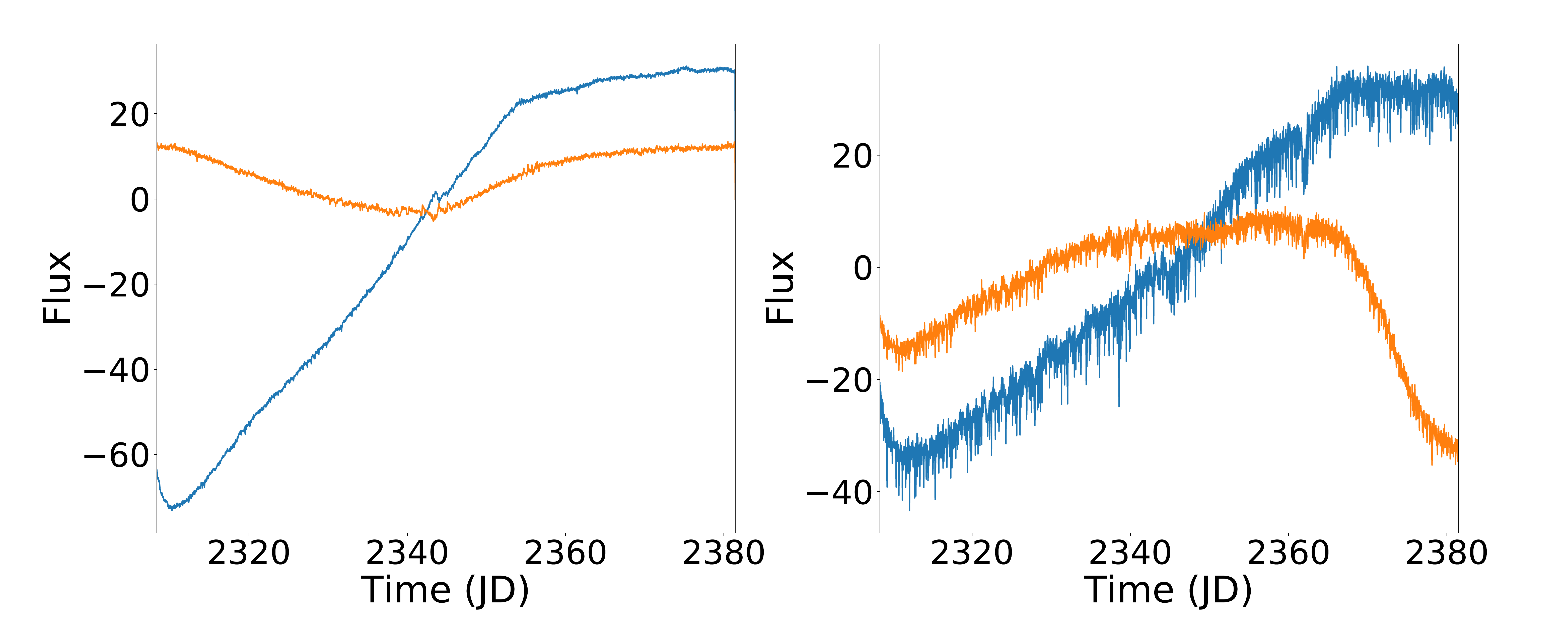}
\caption{Principal components used to correct long-term systematics in the SAP (left) and superstamp (right) datasets as part of the Oxford pipeline.  The primary PCs are in blue, while the secondary PCs are in orange. \label{fig:trends}}
\end{figure*}

Visual inspection of the K2SC-processed light curves reveals significant, residual long-term trends common to many light curves and which appear to be present whatever the extraction and detrending method used.  \citet{2016ApJ...823...16B}, who used the MAST PDC versions of the light curves, already noticed them, and used a Principal Component Analysis (PCA) approach to remove them.  These particular trends are likely due to stellar drift from differential velocity aberration (DVA), which causes photometric aperture losses or variations from sensitivity differences within a pixel.

We also use a PCA approach to remove residual systematics. We normalize each light curve by dividing by its median. We then construct a matrix, where each row is the logarithm of the normalized light curve. (We also experimented with the more standard approach of re-scaling each light curve to have zero mean and unit variance, but found that this gave poorer results). We then use singular value decomposition (SVD) to compute the eigenvectors of the matrix, which are the principal components (PCs); the eigenvalues, which tell us what fraction of the total variance of the dataset is explained by each trend; and a matrix of coefficients relating each light curve to each PC. The first few PCs represent the dominant trends in the data, and these can be removed by subtracting each of them times the corresponding coefficient from each light curve. 

Because M67 falls on two different CCD channels, the two subsets might be expected to show different systematics.  However, we found no significant difference between the sets of PCs extracted from the entire set, or from each output channel in turn. We therefore processed all the SAP stars on the two channels, both cluster members and non-members, together.  On the other hand, there were much more significant differences between the PCs extracted from the SAP and SS sets, as illustrated in Figure~\ref{fig:trends}, so these were processed separately.  The SS PCs appear noisier than the SAP PCs as a result of the relative dominance of the common-mode systematics in each set.  The first, most dominant PC of the SAP set accounts for roughly 84\% of the total variance of the light curves, while the first PC of the SS set only accounts for about 43\%.  The amplitude of the SS PCs are therefore much smaller relative to the noise than the SAP PCs.

If PCA is applied blindly, large-amplitude features in individual light curves can dominate the PCs, skewing the correction and introducing these features in the corrected light curves of other stars. This can be diagnosed easily, however, as the PCs are linear combinations of the light curves, and the linear combination becomes dominated by a single star. Two problematic stars were identified in this way in the SAP set, and excluded from the PC estimation: EPIC 211391083 (an eclipsing binary) and EPIC 211327533 (whose light curve contains a large discontinuity). We also excluded the first 85 cadences ($\sim$1.8\,d) from the light curves before evaluating the PCs, as the systematics were particularly pronounced in that early phase of the campaign, and the PCA correction was of significantly lower quality when they were included (perhaps because the assumption of linearity, which is intrinsic in PCA, was violated more radically in that period).

Usually, a threshold in the fraction of explained variance (i.e. the eigenvalue) associated with each PC is used to decide how many PCs to include in the correction. We experimented with different values for this threshold, and settled on thresholds of 2.5\% and 7.5\% for the SAP and SS sets respectively, which correspond to 2 PCs in both cases, shown in Figure~\ref{fig:trends}.  We tested a range of thresholds for each dataset and found that too high of thresholds only produced one PC, which was not sufficient to explain all of the dominant, common modes seen in the light curves when examined by eye, while too low produced three or more, in which case the extra PCs were superfluous or introduced unwanted features in the data.  A range of thresholds can produce the same 2 PCs, but the ones we chose were the on the lower end of those ranges for both datasets. 

\subsection{The `CfA' pipeline}
In analogy with the Oxford pipeline, the `CfA' pipeline is comprised of the following three segments:
\begin{itemize}
\item Light curve extraction from calibrated K2 target pixel files (as described by \citealt{2014PASP..126..948V} and \citealt{2016ApJS..222...14V})
\item Correction for the 6-hour roll systematics (again as described by \citealt{2014PASP..126..948V} and \citealt{2016ApJS..222...14V})
\item Correction of non-roll systematics using the \emph{Kepler} team's Presearch Data Conditioning (PDC) technique \citep{1538-3873-124-919-1000,2014PASP..126..100S}.
\end{itemize}

\subsubsection{Light curve extraction}
\label{subsubsec:extract2}
Light curve extraction for the CfA pipeline proceeds following the prescription of \citet{2016ApJS..222...14V}, with a few modifications for working on cluster light curves and in crowded regions. Unlike the Oxford pipeline, the CfA pipeline extracts light curves from the calibrated K2 target pixel files for both stars in the superstamp and in individual target postage stamps. We begin by using the World Coordinate System astrometric solution produced by the K2 pipeline to identify the star of interest in either the target postage stamp or the superstamp sub-aperture\footnote{Unlike the Oxford pipeline, we did not splice together the superstamp sub-apertures for our analysis, and instead worked with the 50x50 pixel sub-apertures created by the \emph{Kepler} pipeline. When stars fell near the edge of one of the sub-apertures, we spliced together neighboring sub-apertures to produce the light curves.}. We then laid down 20 different stationary, pixelated photometric apertures over the star in question: ten of these apertures are circular, with radii logarithmically spaced between 1.5 and 13 pixels; the other ten are shaped like the \emph{Kepler} pixel response function (PRF) with different sizes. The PRF-shaped apertures tend to be smaller than the circular apertures. After defining the 20 different apertures, we summed the calibrated, background-subtracted pixels within those apertures at each time stamp to create raw light curves.

\subsubsection{Star-by-star systematics correction}

After producing the raw light curves, we removed the dominant systematic errors on 6-hour timescales introduced by K2's unstable pointing using the method described by \citet{2014PASP..126..948V} and \citet{2016ApJS..222...14V}. This correction works by decorrelating variability due to the roll of the spacecraft (which is correlated with the spacecraft's roll angle)  from astrophysical variability (which is not correlated with the roll angle). Over the course of a K2 campaign, the motion of the spacecraft causes stars to trace out a path back and forth on the detector, which slowly wanders as DVA causes the apparent positions of stars to move over the course of a campaign.  We break up the campaign into a number of shorter time segments, when the motion of the spacecraft can be approximated as purely one-dimensional, and perform our systematics corrections in those divisions individually. We decorrelate the systematics from the astrophysical variability by iteratively dividing away low-frequency variations, determining the relationship between the measured flux and \emph{Kepler}'s roll angle, fitting a piecewise linear function to that relationship (with outlier exclusion to prevent transits, flares, or other short-timescale variability from being mistaken as a spacecraft systematic and removed). Once we have a piecewise linear function which models the roll-angle dependence in the raw light curve, we divide it away, remove any leftover low-frequency variations from the resulting light curve, and repeat the process. After a few iterations, the process converges, and we divide out the best-fit roll-angle variability from the raw light curve to yield a systematics-corrected light curve, with stellar variability and transits preserved. 

After removing systematics from each of the 20 raw light curves produced from the various photometric apertures, we selected a `best' aperture to use in our analysis by determining which aperture produced the light curve with the best photometric precision. In the crowded M67 region, we only allowed the PRF-shaped apertures and circular apertures smaller than 3 pixels in radius to be chosen to prevent additional stars from falling within the photometric aperture. 

\subsubsection{Common-mode systematics correction} \label{subsubsec:cfa_commonmode}
The CfA star-by-star systematic correction above performs very well at removing roll-induced errors in each light curve. 
However, just as with the Oxford light curves, there still exist significant common-mode systematics as a result of focus changes, along with other long-term drifts
due to DVA and other stochastic errors (such as attitude tweaks, safe modes, etc.).  PDC-MAP, on the other hand, was designed to address the
focus-induced systematics that were dominant in the original \emph{Kepler} mission. An ideal K2 pipeline can therefore be
created by applying the CfA correction first followed by the PDC correction.  PDC uses a method somewhat similar to PCA, but it utilizes a Bayesian maximum \textit{a posteriori} (MAP) approach, where a subset of highly correlated and quiet
stars is used to generate a co-trending basis vector set, which is in turn used to establish a range of robust, `reasonable' fit parameters. These parameters are then used to generate a Bayesian prior and a Bayesian posterior
probability distribution function (PDF) which, when maximized, finds the best fit that simultaneously removes systematic
effects while reducing the signal distortion and noise injection that commonly afflicts simple least-squares
fitting.  A numerical and empirical approach is taken where the Bayesian prior PDFs are generated from fits to the
light-curve distributions themselves.

PDC has two modes of operation: `Single-Scale' (ssMAP) and `Multi-Scale' (msMAP). The former, ssMAP
\citep{1538-3873-124-919-1000}, performs the MAP correction in a single band-pass. This has the advantage of instilling
a strong regularization on the basis vector fit coefficients, thereby minimizing the removal of stellar signals.  However, this
is at the expense of more bias and less systematic error removal. msMAP \citep{2014PASP..126..100S}, on the other hand,
utilizes an over-complete, discrete wavelet transform, dividing each light curve into multiple channels, or bands. The
light curves in each band are then corrected separately, allowing for a better separation of characteristic signals and
improved removal of the systematic errors, but at the expense of more stellar signal removal, especially at longer
periods.  Both methods have their advantages, but generally speaking, msMAP performs better at preserving and cleaning light
curves at transit time scales, and ssMAP performs better at preserving long period signals. It is therefore expected that
ssMAP will perform better for the studies carried out in this paper.

\subsection{Scatter comparison} \label{subsec:p2p}

\begin{figure*}
\includegraphics[width=1\linewidth]{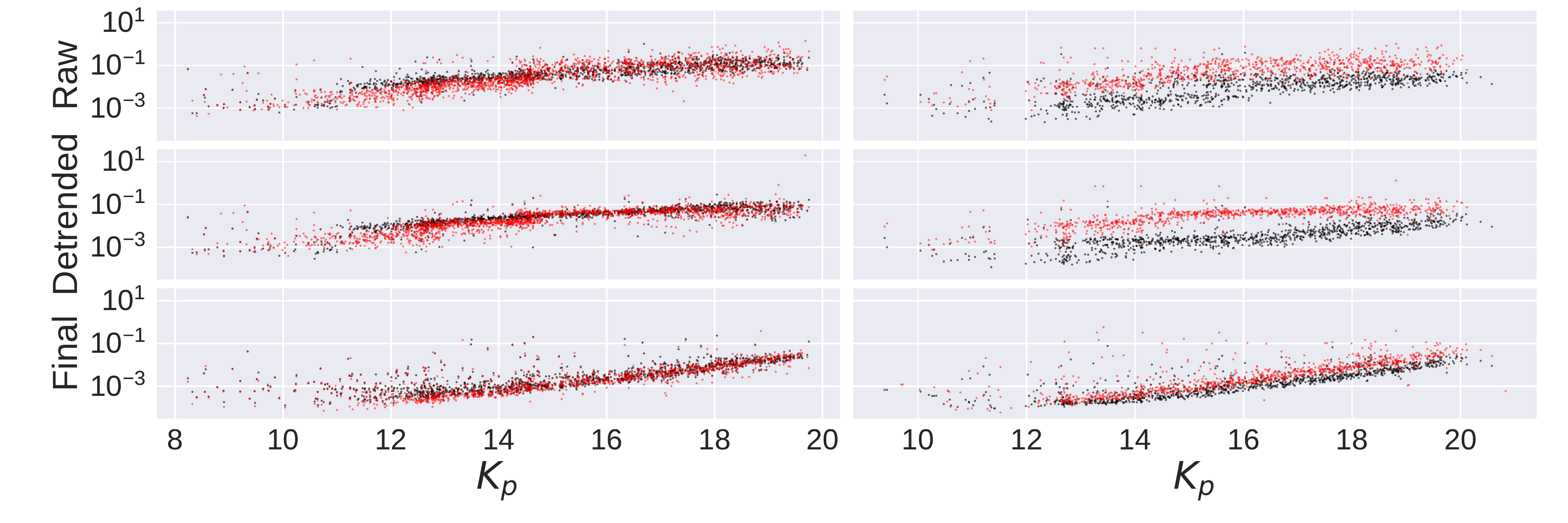}
\caption{Scatter versus $K_{p}$ magnitude for the SAP (left column) and SS (right column) samples. The results of the Oxford and CfA pipelines are shown in black and red, respectively. From top to bottom, we plot the scatter in the raw light curves, after star-by-star systematics correction, and after common-mode systematics correction. \label{fig:p2pscatter}}
\end{figure*}

We here compute the scatter of the M67 SAP and SS stars as a function of \emph{Kepler} magnitude at each step of the Oxford and CfA pipelines in order to conduct a preliminary assessment of pipeline performance prior to the injection tests.  We use a variation of the median of absolute deviations (MAD) \citep{1983ured.book.....H} scatter, estimated in the following manner:

\begin{equation} \label{eq:p2p}
	\textrm{MAD} = 1.48\textrm{median}(|f_{i} - m_{f}|)
\end{equation}

\noindent where $f_{i}$ is the flux at each cadence $i$, and $m_{f}$ is the median flux.  We use the median as opposed to the mean because it is less influenced by outliers.  Technically, the median term itself is the MAD scatter; we include a factor of 1.48 as a scaling factor that, under the assumption of a Gaussian distribution of $f$, makes the scatter equivalent to the standard deviation.

The scatter is given in Figure \ref{fig:p2pscatter}.  The Oxford pipeline results are shown in black and the CfA, using single-scale PDC-MAP, in red. From top to bottom, the figure shows the scatter for the raw light curves, after star-by-star systematics correction, and after common-mode systematics removal, with the SAP sample on the left and the SS sample on the right. In the SAP sample, the raw scatter produced by the two pipelines are fairly similar, though the CfA pipeline performs slightly better. For both pipelines, the star-by-star systematic correction reduces the scatter somewhat over the whole magnitude range and forms tighter relationships, but the final common-mode systematic correction has the most drastic effect, leading to scatter values of a few tens of ppm at the bright end. The final scatter obtained by the CfA pipeline is slightly lower than that obtained by the Oxford pipeline over the entire magnitude range.  The final median scatter at $K_{p}=12$ for the Oxford and CfA SAP samples are 544\,ppm and 321\,ppm, respectively.

By contrast, the raw outputs of the two pipelines are very different for the SS sample. The extraction step used by the Oxford pipeline, which involves moving apertures, already lessens the pointing-related systematics in the raw light curves of fainter stars.  The effect of the star-by-star systematic correction on the scatter is then relatively minor for both pipelines, but the common-mode systematic correction is particularly effective over the whole magnitude range for the CfA pipeline, and for bright stars in the Oxford pipeline.  The final results of the two pipelines are fairly similar, though the CfA pipeline performs slightly better for brighter stars (not to the extent of the SAP sample, however). Among the fainter stars, there are several instances where the CfA pipeline leads to final scatters of order 1\%, whereas the corresponding scatter for the Oxford pipeline is much lower.  In these cases, the Oxford pipeline generally tends to correct the systematics better.  The final median SS scatter at $K_{p}=12$ is 203\,ppm for the Oxford pipeline and 275\,ppm for the CfA pipeline. 

The final scatter distributions are very similar for the SAP and SS samples, as well as for the two pipelines.  It is worth noting, however, that the scatter between the CfA SAP and SS datasets are closer to each other than their counterparts in the Oxford pipeline.  This is most likely due to the fact that the CfA light curves are all extracted using the same technique, while the two Oxford datasets are extracted via different methods, and so have different starting points with respect to noise levels.  Regardless, these numbers should give us increased confidence that the light curves we are using, from both pipelines, are relatively robust, and there is no obvious effect in either sample that is not handled reasonably well by either pipeline.  

\section{Injection Tests} \label{sec:tests}

We seek to quantify the limits of our ability to measure stellar rotation periods in K2 M67 light curves. To do this, we need to simulate realistic light curves that share the same time sampling and noise properties as those we wish to analyze, but which also contain signals of known period and amplitude (as well as any other parameter that might affect the detectability of the signal). In the absence of a detailed generative model for the various noise sources and systematics in K2 data, we are not able to simulate realistic light curves from scratch.  Instead, we inject known rotation signals into the raw versions of the observed SAP and SS light curves (i.e. just after extraction).  This introduces a few problems, the most obvious being that some light curves may already contain strong astrophysical variability, to which we later return.  We then separately apply the detrending steps (pointing-related and common-mode systematics correction) of the Oxford and CfA pipelines, before attempting to recover the periodic signals.  This enables us to assess the extent to which the noise present in the data, as well as the pre-processing applied to reduce this noise, affects the detectability of the periods.

As described in Section~\ref{sec:intro}, this paper deals only with sinusoidal injected signals, but the tests are designed to be comprehensive, in that every signal is injected into every light curve. A smaller set of tests involving more realistic signals and comparing different period search methods will form the subject of a later paper.

\subsection{The stellar samples} \label{subsec:star_samples}

For the purposes of the injection tests, we define the following 3 stellar samples:
\begin{itemize}
    \item the SAP sample consists of the 1877 stars observed with individual postage stamps in channels 6.1 and 6.2, 1319 which match to our master catalog and 236 of which are confirmed members from G15 
    \item the SS sample consists of the 976 stars included in the superstamp and with matches in the master catalog, 359 of which are confirmed members from G15. 75 of the SS overlap with the SAP
    \item in addition, we define a hot star sample consisting of 89 main sequence stars with spectral type A through F located on CCD modules 6, 11, and 16 with effective temperatures ranging from 6300\,K to 10715\,K (the majority falling between 6300 and 7300\,K), $B-V$ values of -1.0 to 0.45, and $\textrm{log}\,g$ values greater than 4.0.  This information was all acquired from MAST.  These stars were chosen because they lack an outer convection zone, or have only a very thin one, and so are not able to support a large-scale magnetic field and so do not spin-down quickly \citep{1962AnAp...25...18S}.  Thus, if they do display periodic variability, it is expected to occur on timescales of $\sim2$\,d or less, whether it is due to pulsations (many of these stars lie in the classical instability strip) or rotation (as these hot stars are rapid rotators) \citep{2013A&A...557L..10N}. Therefore, their light curves are expected to contain only noise and systematics, at least on timescales longer than $\sim$2 -- 7\,d, which makes them ideal targets for injection tests. 22 of the hot stars overlap with the SAP set
\end{itemize}

\subsection{The injected signals} \label{subsubsec:injsig}
We inject sinusoidal signals by directly multiplying the following into the raw light curves of each test sample:

\begin{equation} \label{eq:injsig}
	f_{\rm inj} = \frac{a_{\rm inj}}{2}\sin(\frac{2\pi}{P_{\rm inj}}t+\phi_{\rm inj}) + 1 
\end{equation}

\noindent where $f_{\rm inj}$ is the injected signal, $a_{\rm inj}$ is the injected amplitude, $P_{\rm inj}$ is the injected period, $\phi_{\rm inj}$ is the injected phase, and $t$ is time in days.  We vary $P_{\rm inj}$ from 5 to 35\,d in intervals of 5\,d, leading to seven period values.  We also vary $a_{\rm inj}$, as follows: 0.05\%, 0.10\%, 0.30\%, 0.50\%, 1.00\%, and 3.00\%.  Each light curve, then, is injected with 42 different combinations of periods and amplitudes, leading to over 120\,000 injections in total for each pipeline to process.  The phase, $\phi_{\rm inj}$, is selected at random for each injection, but we keep track of its value. Where we have extracted the SS light curves using multiple aperture sizes from the Oxford pipeline, we inject the signals into the version which minimizes the scatter for the corresponding, non-injected light curve.

\subsection{The Detection Algorithm} \label{subsubsec:detalg}
Because these injection tests involve only sinusoidal signals, we use a modified version of the widely-used Lomb-Scargle (LS) periodogram \cite{1982ApJ...263..835S} to recover them. The LS periodogram is essentially equivalent to least-squares fitting of sinusoidal signals \citep{2006MNRAS.370..954I}, and is thus in principle optimal to recover stable sinusoidal signals in the presence of white Gaussian noise of known variance. 

Real stellar rotation signals are not strictly sinusoidal; they contain significant power at harmonics of the true period due to the scattered distribution of the active regions on the stellar surface, evolve over time with the active region evolution, and may contain signals at a range of periods due to differential rotation.  For this reason, rotation  period searches in \emph{Kepler} and K2 data often use other period search methods which do not depend on the assumption of sinusoidality or strict periodicity \citep{2013MNRAS.432.1203M,2017arXiv170605459A}. These methods have been shown to outperform approaches based on least-squares sine-fitting in real and simulated datasets \citet{2015MNRAS.450.3211A}.  However, if we know the signal of interest is sinusoidal, then a sine-fitting approach is likely to do at least as well as these alternative, more flexible methods. 

The model we are considering is of the form:
\begin{equation} \label{eq:pmodel}
	m(t) = m_{dc} + \alpha\sin(\omega t + \phi) 
\end{equation}
\noindent where $m_{\rm dc}$ is the light curve's vertical offset, $\alpha$ is the amplitude, $\phi$ is the phase, and $\omega$ is the angular frequency.  For each value of $\omega$ (or period, $P=2\pi/\omega$), this model can be expressed as a linear basis model with three basis functions: a constant, a sine term, and a cosine term, both with period $P$. For each trial period we can solve for the values of $m_{\rm dc}$, $\alpha$, and $\phi$ that minimize the sum of the squared residuals, or $\chi^2$ (in space data, the relative measurement uncertainties can be treated as approximately constant for a given star, so the two are equivalent). We evaluate the relative reduction in $\chi^2$ with respect to a constant model, $S = (\chi^2(0)-\chi^2(P))/\chi^2(0)$, as a function of period $P$, to construct a periodogram. For this study, we search for 490 periods ranging from 2 to 100 days, using an evenly-spaced grid in frequency space.

\subsubsection{Periodogram normalization} \label{subsubsec:per_norm}

\begin{figure}
\plotone{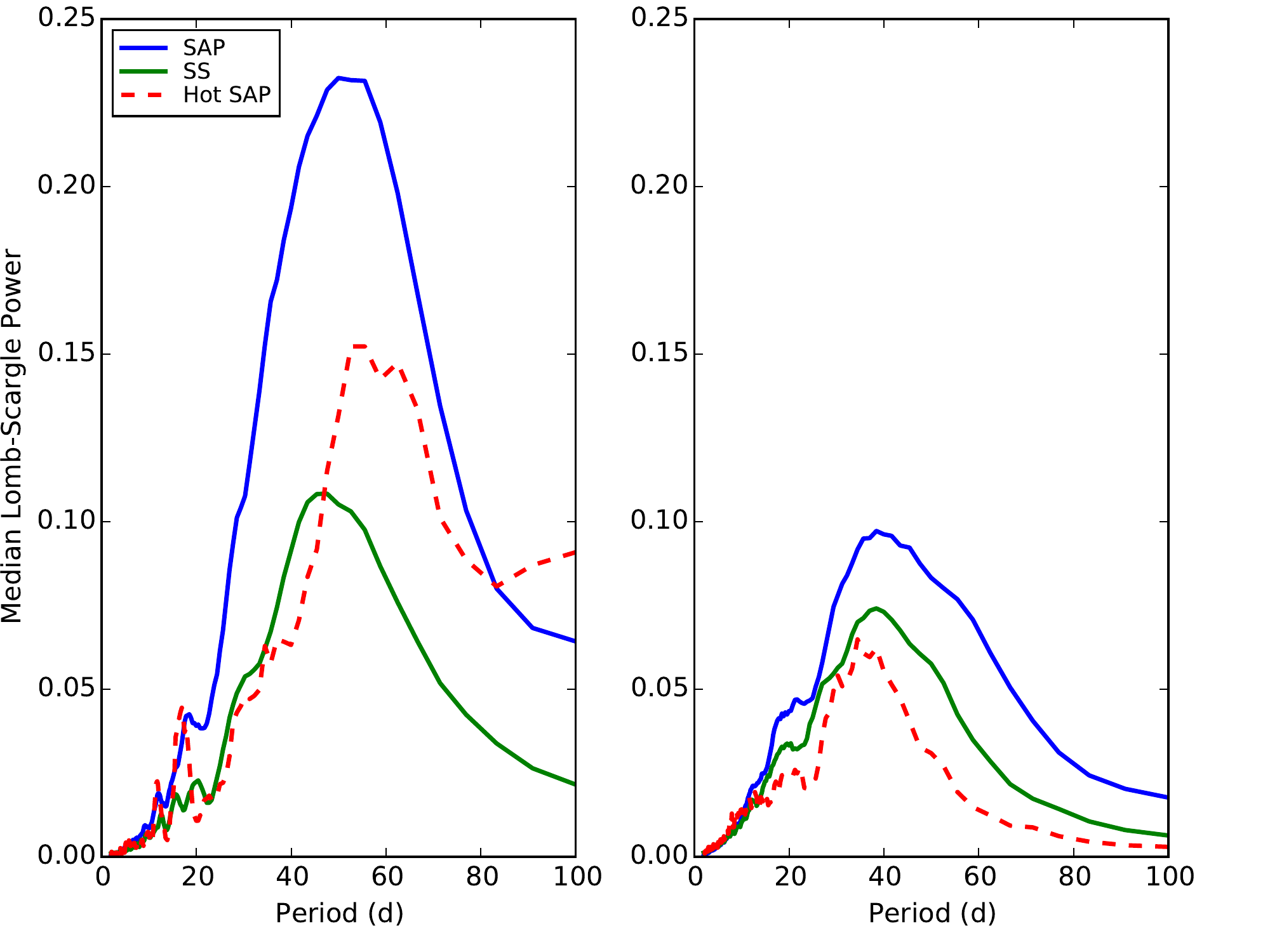}
\caption{Median Lomb-Scargle periodogram for the non-injected SS (blue) and SAP (green) datasets for both pipelines. The non-injected hot star dataset for each pipeline separated from the SAP dataset is shown with red dashed lines. The Oxford pipeline is on the left, and the CfA pipeline is on the right. The presence of broad peaks around 30\,d and longer in the hot star sample highlights the presence of long-term, non-astrophysical signals in the Campaign 5 light curves \label{fig:normal}}
\end{figure}

If the light curves consisted simply of a stable sinusoidal signal plus white Gaussian noise, locating the peak from the Lomb-Scargle periodogram would give us the best-fit period, and evaluating the significance of the detection would be relatively straightforward \citep[see e.g.][]{1986ApJ...302..757H}. However, this is not the case, even after our light curves are subjected to the systematics-correction steps of our two pipelines. The residual systematics, which remain in the light curves despite our best efforts, often lead to broad peaks in the range 30 -- 60\,d, which can overwhelm the peak due to the injected signal. To address this, we perform a collective normalization of the periodograms before searching for the period. 

\begin{figure}
\plotone{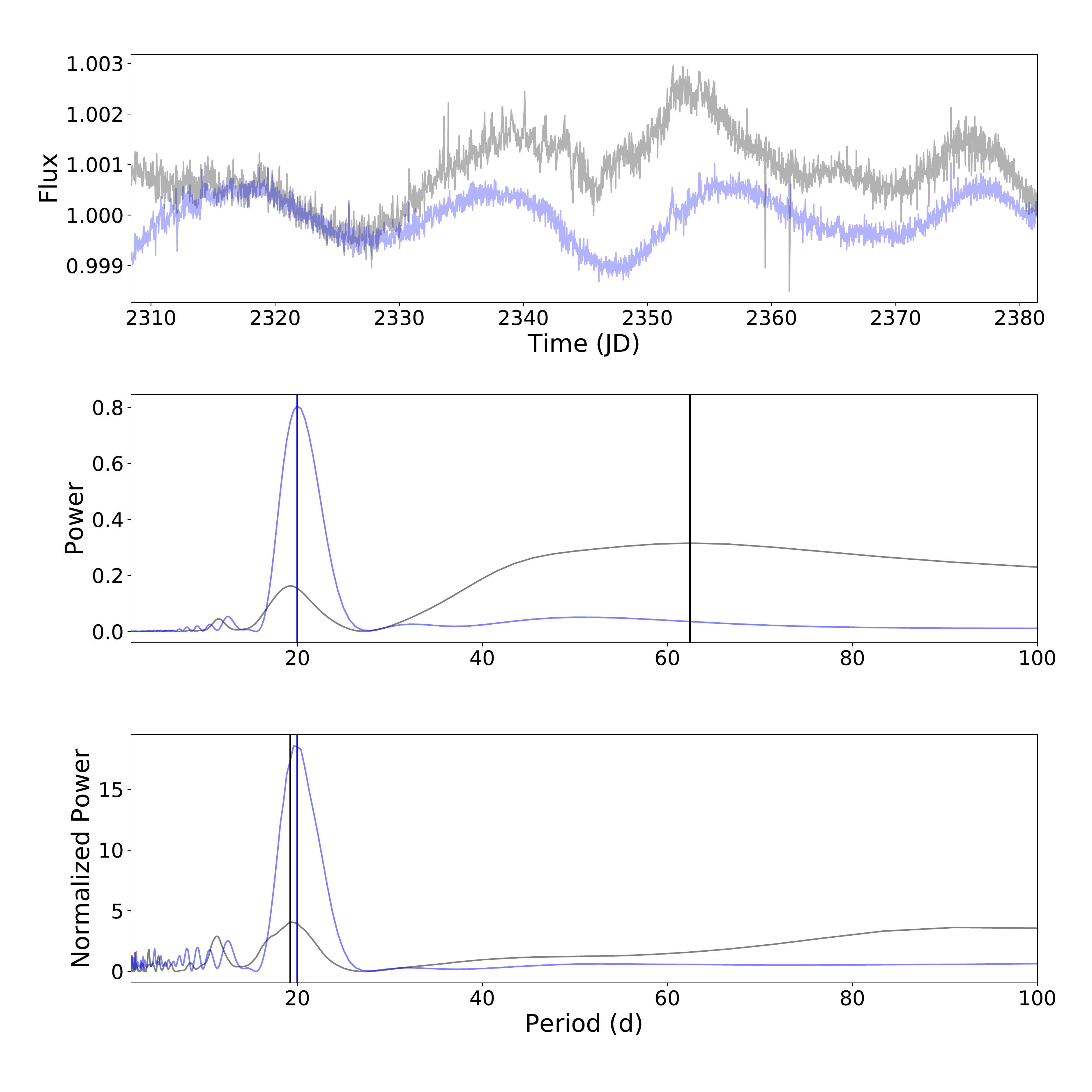}
\caption{Example of the effect of periodogram normalization using EPIC 211355490 injected with a period of 20\,d and amplitude of 0.1\%.  The top panel shows the systematics-corrected flux from the Oxford (gray) and CfA (blue) pipelines.  The middle panel gives the original periodograms for both pipelines, while the bottom panel shows the normalized periodograms. The periods found in each periodogram are indicated by the vertical lines, with the black lines depicting the Oxford periods and the blue lines identifying the CfA periods.  The CfA period in both cases is 20.0\,d, while the Oxford period improves from about 62\,d to 19.2\,d. \label{fig:renorm_example}}
\end{figure}

To normalize the periodograms, we compute the median periodogram as a function of trial period for stars which we expect to share the same systematic noise properties (i.e. separately for the SAP and SS sets, and for the Oxford and CfA pipelines). Each star's periodogram is then divided by the median periodogram, thereby suppressing the peaks, or excess power, that is seen in many light curves.  The normalized periodogram is thus given by $S'(P) = S(P)/\hat{S}(P)$, where $\hat{S}(P)$ is the median periodogram value for period $P$.  The median periodograms are evaluated from the non-injected versions of the light curves to avoid the induced power at fixed periods from their injected counterparts, and they can be seen in Figure~\ref{fig:normal}, with the Oxford SS and SAP median periodograms on the left and the CfA on the right.  The SAP median periodograms for both datasets include the hot stars, as they were processed in both cases using the same techniques.  However, Figure~\ref{fig:normal} also shows the hot star median periodograms separate from the rest of the SAP set for both pipelines, given by red dashed lines.  This emphasizes that the broad peaks in both pipelines at long periods are most likely not a result of true astrophysical signals, as we would not expect periods much longer than about 5\,d in the hot star samples.  We also note that the median periodogram for the SAP set from the Oxford pipeline is higher overall and peaks at longer periods than the SS sample, hinting that the former likely contains systematics with larger amplitudes on greater timescales. This is consistent with the relative amplitudes of the PCA trends compared to noise for the two samples, which are smaller for the SS (see Figure~\ref{fig:trends}.  It also means that long-period signals will be more heavily suppressed by normalization in the SAP than in the SS sample.

Figure \ref{fig:renorm_example} illustrates the effect of the periodogram normalization for an individual star from the SAP set. The top panel shows the light curve of EPIC 211355490 injected with a period of 20\,d and amplitude of 0.1\%, fully processed with both the Oxford (in gray) and the CfA (in blue) pipelines.  The middle panel depicts the original periodograms prior to normalization for both pipelines, while the bottom panel gives the normalized versions.  The respective periods are indicated by vertical lines.  There is a relatively strong, long-period signal in the Oxford pipeline that is not present in the CfA (unsurprising given the differences in Figure~\ref{fig:normal}), and this is what the Lomb-Scargle finds in the former, settling on a period of 62\,d.  On the other hand, the normalized Oxford periodogram has a peak at 19.2\,d, which closely matches the injected period.  The normalization of the periodogram thus allowed us to find the `true' period of Oxford version of this light curve when otherwise it would have been lost by the systematic, long-term trend.  The CfA, however, appears to have done a better job of cleaning up systematics in the light curve, and found peaks of 20.0\,d in both the original and normalized periodograms.

\subsubsection{Detection threshold}

Once we have identified the peak in each normalized periodogram, we must decide whether it is significant.  After testing a number of different possible schemes, we opted for the following detection threshold definition:
\begin{equation} \label{eq:detection}
	S'_{\rm T} = {\rm MAX}(S'_{90} \times C, S'_{\rm min})
\end{equation}
\noindent where $S'_{T}$ is the threshold value and $S'_{90}$ is the $90^{\rm th}$ percentile of the normalized peridogram for a given light curve. $C$ and $S_{\rm min}$ are tuning parameters which we can vary to alter the relative and absolute components of the threshold, respectively. For example, if $S'_{90} = 4$, and the value for $C$ is set at 3, then the maximum power of the normalized periodogram must be at 12 or greater for the associated period measurement to be considered a detection.  $S_{\rm min}$ comes into play when $S'_{90} \times C$ is so low that most peaks would qualify as a detection; it ensures there's a minimum value for the threshold.  Both $C$ and $S_{\rm min}$ can then be varied to test their effect on the completeness and reliability of the period search, which we discuss in Section~\ref{subsubsec:vary_thresh}.

If the highest peak in the normalized periodogram passes the threshold defined in Eq.~(\ref{eq:detection}), then we have a detection, and we record the corresponding period, as well as the best-fit amplitude and phase. Otherwise, a result of `no detection' is recorded.

\subsection{Evaluating the results}

To evaluate the results of the injection tests, we need to quantify how sensitive the period search is, i.e. what fraction of the injected signals are successfully recovered, but also how trustworthy the detections are, i.e. what fraction of the detections are `valid', meaning that the measured period and phase are within some tolerance of the injected values. 

\subsubsection{Valid detections, completeness and reliability}
Here we define several key terms in understanding the statistics from this study:

\begin{itemize}
    \item Validity: We consider a detection \emph{valid} if the recorded period and phase are within 20\% of the injected values.
    \item Completeness: For a given range of \emph{injected} period and amplitude, we define the \emph{completeness} as the fraction of injected signals which led to a valid detection. We also calculate a \emph{threshold error} statistic, which records the fraction of cases where the period and phase corresponding to the highest peak in the normalized periodogram were within the valid range, but the peak was not high enough to pass the detection threshold.
    \item Reliability: For a given range of \emph{recorded}, or measured, period and amplitude, we define the \emph{reliability} as the fraction of recorded detections which were valid. 
\end{itemize}

Another way to understand the definitions of completeness and reliability given above is to look ahead at Figure~\ref{fig:detvinj1}, which shows (for the hot star sample) the detected versus injected periods for different injected amplitudes. To evaluate completeness, we consider a vertical bin in one of these diagrams around one of the injected periods. The completeness is given by the number of detections (colored points) in that bin which were valid, divided by the total points in the bin. To be valid, a detection must lie within the gray shaded area, which indicates a $\le 20\%$ period error. In addition, a valid detection must also have a $\le 20\%$ phase error (not shown on the figure). We can think of reliability as being similar, but considering a horizontal rather than vertical bin in the same kind of diagram. (This is not quite correct, since reliability is computed for a given detected period and amplitude, whereas the figure shows the injections split by injected amplitude.)

\subsubsection{Calculating Uncertainties}
To estimate the uncertainties, we assume a Poisson distribution with respect to both completeness and reliability and define the uncertainty as:
\begin{equation} \label{eq:uncertain}
	\sigma = \pm\frac{1}{\sqrt{N}}
\end{equation}
\noindent where $N$ is the number of injections per injected period and amplitude bin (i.e. no. of injected amplitudes $\times$ no. of injected periods $\times$ no. of stars in each test sample) in the case of completeness. For reliability, $N$ is the total number of detections in each detected period and measured amplitude range.  Therefore, for each test set within each pipeline, the individual completeness values will have the same uncertainty while the reliability will vary from bin to bin.  We recognize this is a little simplistic, but the more sophisticated approach would require repeating the injections many times, which would be far too time-consuming.  For our purposes, this is a good first approximation.

\subsubsection{Intrinsically variable stars}
We injected signals into real K2 light curves in order to faithfully reproduce the noise properties of actual M67 light curves. In doing this, we are implicitly assuming that most of them do not already contain a detectable intrinsic periodic signal. However, of course, some of them do. In such cases, a detection might be caused by the intrinsic rather than the injected signal. This can lead to detections which appear `invalid', but are in fact correct, biasing the completeness and reliability statistics.

To avoid this, we run the period search on the non-injected versions of the light curves.  If a detection occurs in any of these, the star is labelled as intrinsically variable, and is excluded from the completeness and reliability calculations. This means that only a fraction of the total number of injections we performed is actually used in the final results. 

This is not entirely satisfactory, since in reality we do not know whether the signal detected in the light curves without injections were indeed due to the intrinsic variability of the star or to systematics. Furthermore, the number of stars excluded as `variables' depends on the detection threshold used. However, we were unable to devise a more satisfactory solution. We note that, for reasonable threshold values, the variable stars represent a relatively small minority of the test light curves.

\subsubsection{Varying the threshold} \label{subsubsec:vary_thresh}

\begin{figure*}
\plotone{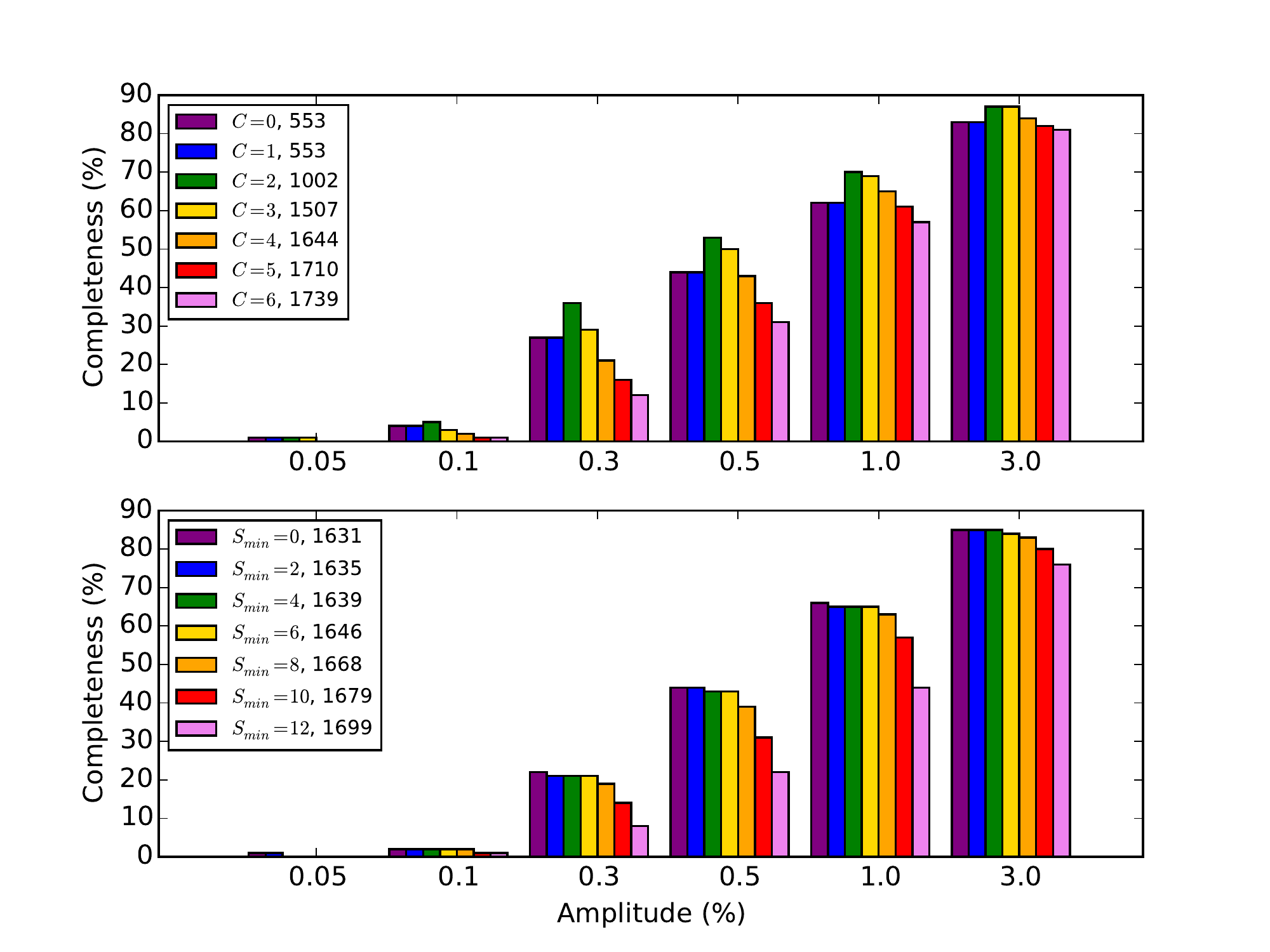}
\caption{Comparison of completeness values for an injected period of 25d at all injected amplitudes for different values of $C$ (top row) and $S_{min}$ (bottom row). For the top row, $S_{min}$ was fixed at 5.  For the bottom row, $C$ was fixed at 4.  The legends are also marked with the number of stars in each injected period and amplitude bin for each value of $C$ and $S_{min}$.  It is worth noting that by $S_{min}=12$, all sensitivity at periods greater than 25\,d was gone. We used the Oxford pipeline for these tests. \label{fig:comp_compare}}
\end{figure*}

\begin{figure*}
\plotone{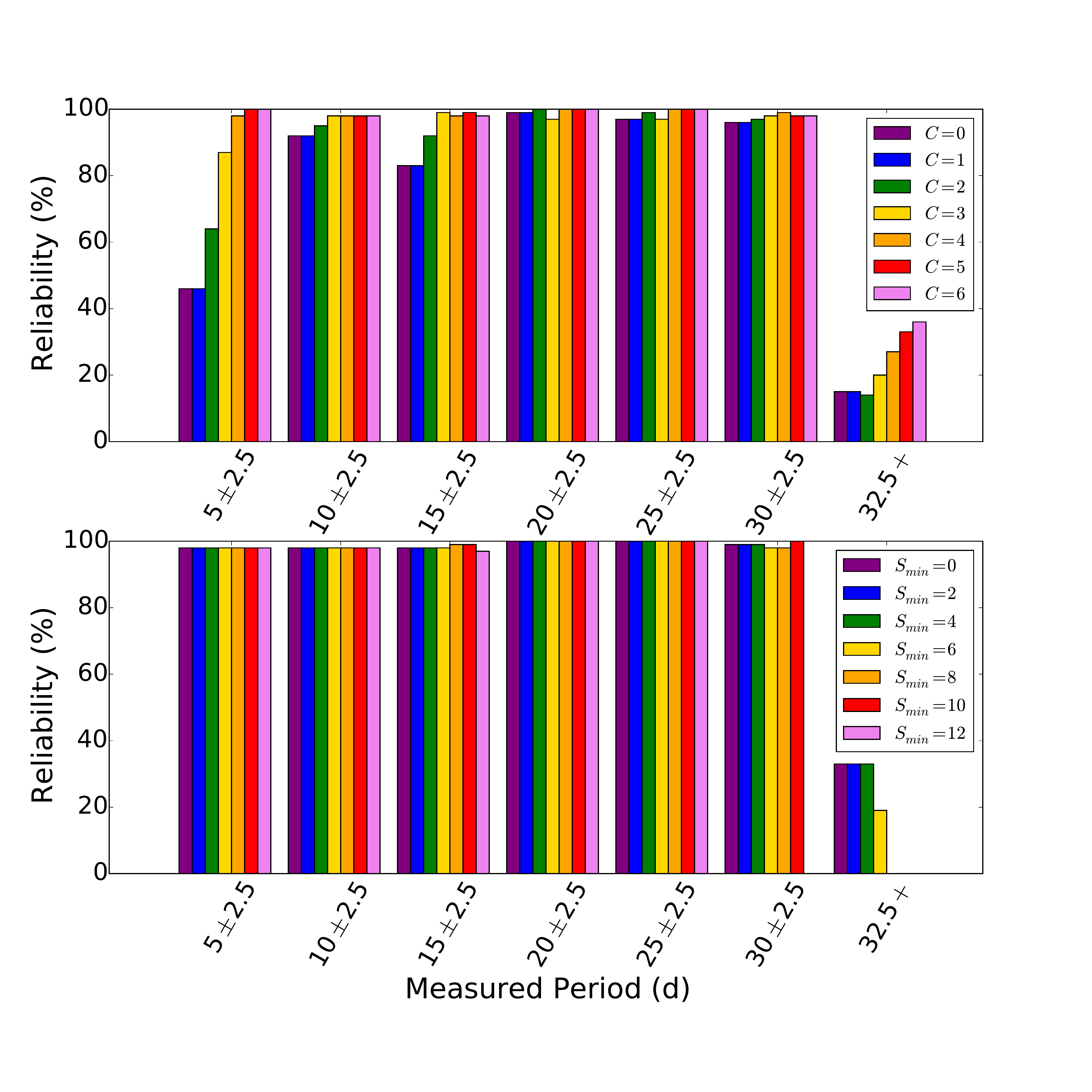}
\caption{Comparison of reliability values for a measured amplitude between 0.25\% and 0.45\% across all detected periods for different values of $C$ (top row) and $S_{min}$ (bottom row).  For the top row, $S_{min}$ was fixed at 5.  For the bottom row, we fixed $C$ at 4.  It is worth noting that we lost sensitivity for detections greater than 32.5\,d at $S_{min}=8$, and we lost sensitivity for detections greater than 27.5\,d at $S_{min}=12$.  We used the Oxford pipeline for these tests. \label{fig:rel_compare}}
\end{figure*}

The detection threshold obviously has an effect on the results of this study.  Lower thresholds increase completeness, but also lead to a drop in reliability in the same bins.  If measured rotation periods are to be compared with theoretical models, it is preferable to prioritize high reliability over slight gains in completeness --- so long as the completeness itself is well-measured to account for missed detections.  We therefore recorded the normalized periodogram peak $S'$; the corresponding period, phase, and amplitude; and the $90^{\rm th}$ percentile of the normalized periodogram, $S'_{90}$, for each injected signal in each light curve. It was then trivial to vary the value of $C$ and $S_{\rm min}$ in \ref{eq:detection} and examine the impact that this had on completeness, reliability, and the `variables' excluded from the statistics.  

We experimented with $C=0$ to $C=6$ while holding $S_{\rm min}$ at 5, and varying $S_{\rm min}$ from 0 to 12 while holding $C$ at 4, in order to test the effect that each parameter has on completeness and reliability.  We used the Oxford pipeline to do these two tests.  A sample of the results of these tests can be seen in Figures~\ref{fig:comp_compare} and \ref{fig:rel_compare}.  Figure~\ref{fig:comp_compare} shows the effect on completeness for changing both tuning parameters at an injected period of 25\,d across all injected amplitudes.  The top row of each figure shows the effect of changing $C$, while the bottom row shows the effect of changing $S_{\rm min}$.  In both panels, we print the number of stars in each injected period and injected amplitude bin.  We can see that at $C=0$ and $C=1$, the results are dominated by the minimum threshold value, but they have relatively few stars per bin. Completeness then peaks at $C=2$ before starting to decline again. There is also an increase of $\sim$500 stars per bin at $C=2$, followed by another 500 stars at $C=3$.  Beyond $C=3$, the number of stars per bin still grows, but less rapidly.  For $S_{\rm min}$, the completeness essentially stays the same until about $S_{\rm min}=8$. By $S_{\rm min}=12$, the completeness drops to about half the values at $S_{\rm min}=0$ at amplitudes of 0.5\% and less.  Not visible in Figure~\ref{fig:comp_compare} is the fact that all sensitivity at periods greater than 25\,d, at any amplitude, is lost.  The number of stars per bin increases with $S_{\rm min}$, but very slightly.  Thus, as expected, $C$ has a much greater effect on the number of stars per bin and, more importantly, completeness, than $S_{\rm min}$, as long as $S_{\rm min}$ is not too high (in the case of completeness).

Likewise, in Figure~\ref{fig:rel_compare}, we can see an illustration of how $C$ and $S_{\rm min}$ affect reliability across the ranges of measured periods in this study.  Here we have fixed the measured amplitude to the range of 0.25\% to 0.45\%, just above the solar range.  Again, the top row shows the effect of changing $C$.  As expected, reliability generally increases as $C$ increases, noticeably so until about $C=4$. The effect of changing $S_{\rm min}$, however, is largely negligible except at the long period ranges.  There, it decreases until $S_{\rm min}=8$.  Here the reliability is zero, but with very few detections, and beyond, the number of detections drops to zero for periods longer than about $\sim$30\,d.  

Figures~\ref{fig:comp_compare} and \ref{fig:rel_compare} demonstrate the clear trade-off between completeness and reliability.  Using large values for $C$ and $S_{\rm min}$, i.e. a stringent detection threshold, leads to low completeness and large threshold error, but excellent reliability. A high detection threshold also means that fewer stars are marked as intrinsically variable, so that more injected light curves are used to compute the final statistics.  As the threshold is gradually lowered, completeness increases and threshold error decreases, at the cost of reduced reliability.  The number of variable stars also grows, and for very low detection thresholds almost all stars are excluded, leaving relatively few in the calculation of the results.  Therefore, we settled on $C=4$ and $S_{\rm min}=4$ as the best compromise.  When using these values, 238 of the 1877 stars in the Oxford SAP set were marked as intrinsically variable, as were 120 of the 976 stars in the SS set and 23 of the 89 hot stars, leaving 1639, 856, and 66 stars in each set, respectively.  Using the same tuning parameters for the CfA pipeline, we are left with 1587 SAP stars, 848 SS stars, and 70 hot stars. 

\section{Results} \label{sec:results}
We restrict ourselves to a presenting the results in this section and defer a more detailed discussion to Section~\ref{sec:discussion}.  The completeness and reliablity results for the CfA pipeline come from the single scale PDC-MAP, or `PDC-ssMAP.'  Across the board, PDC-ssMAP performed better, for our purposes, than PDC-msMAP, as expected (see Section~\ref{subsubsec:cfa_commonmode}).  

\subsection{Recovered versus injected periods}

\begin{figure*}
\plotone{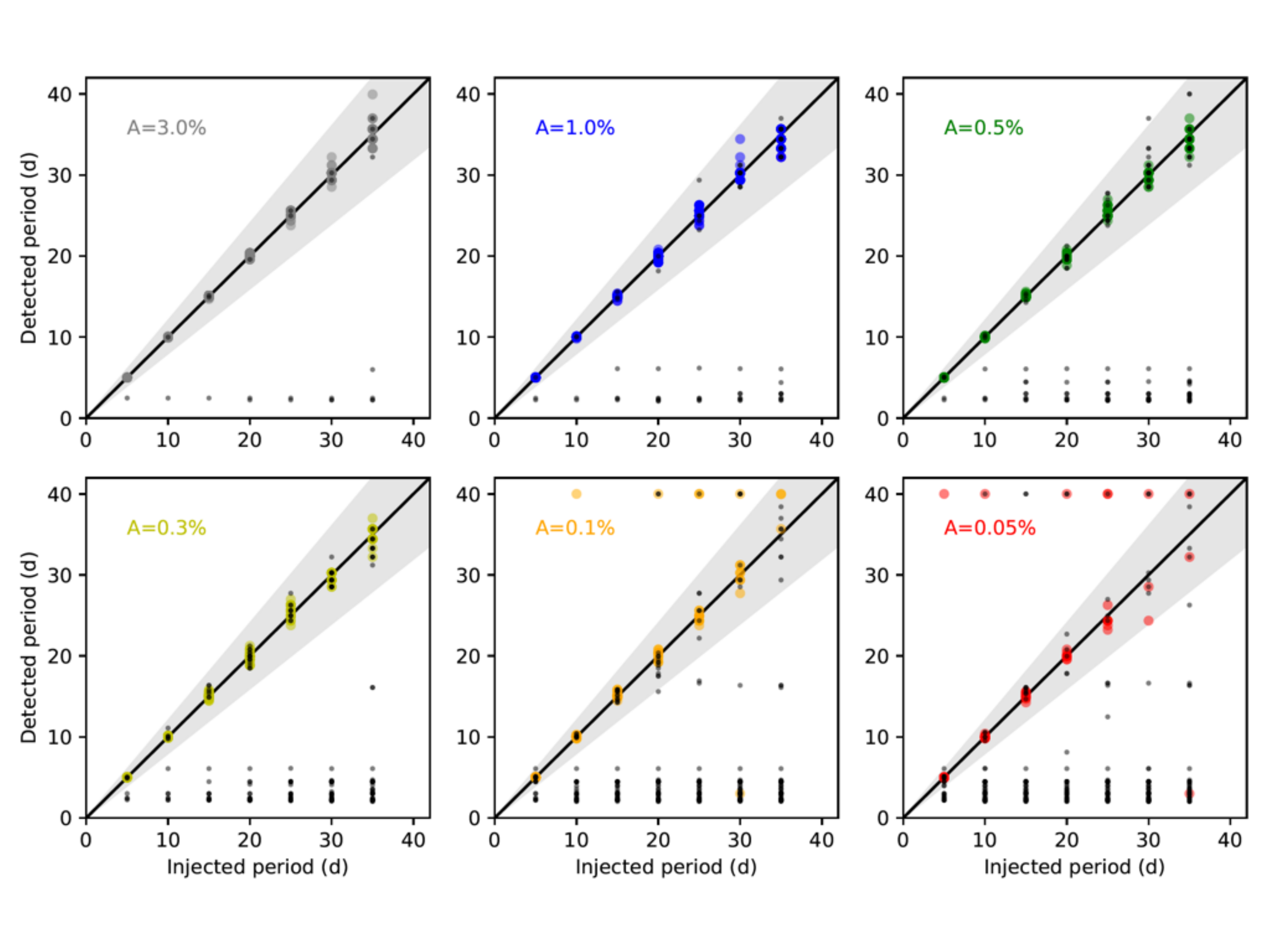}
\caption{Detected versus injected period at each of the injected amplitudes for the hot star sample processed with the Oxford pipeline. Colored points represent cases which passed our detection threshold; those which did not are shown as smaller black points. The shaded grey area shows the region where the injected and detected periods are within 20\% of each other. \label{fig:detvinj1}}
\end{figure*}

\begin{figure*}
\plotone{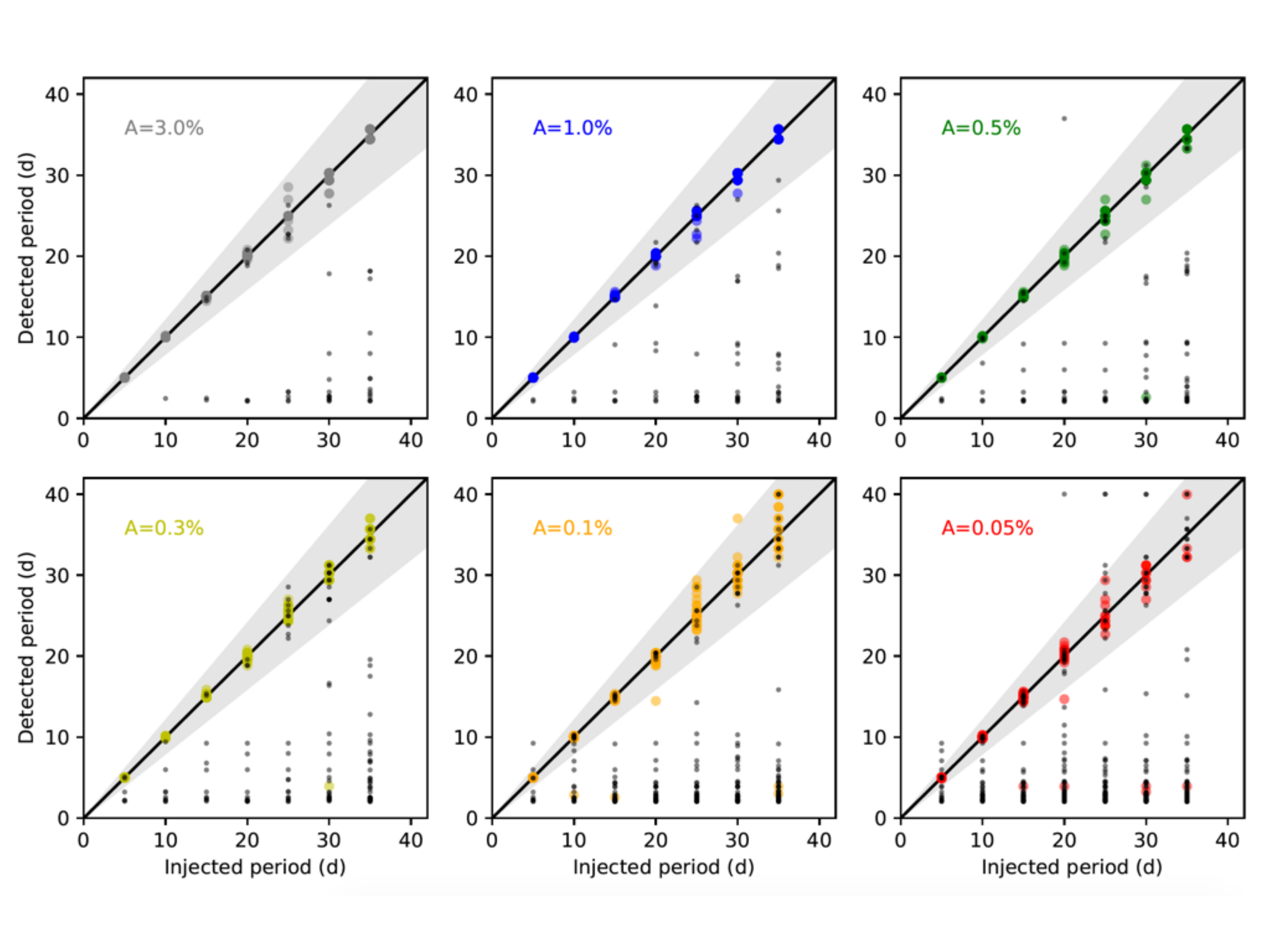}
\caption{Same as Figure~\ref{fig:detvinj1}, but for the hot star sample processed with the CfA pipeline. \label{fig:av_detvinj1}}
\end{figure*}

\begin{figure*}
\plotone{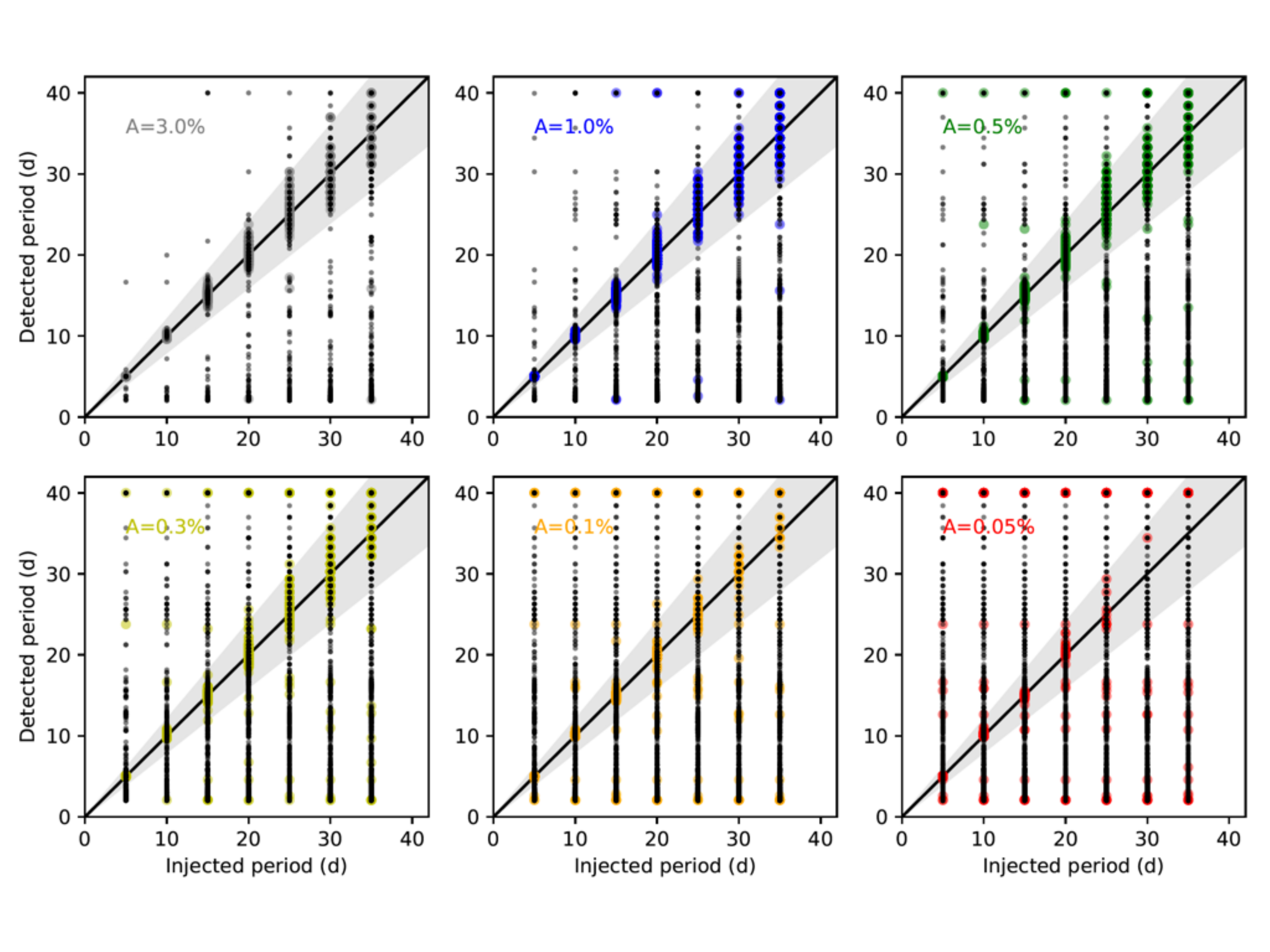}
\caption{Same as Figure~\ref{fig:detvinj1}, but for the SAP sample processed with the Oxford pipeline. \label{fig:detvinj2}}
\end{figure*}

\begin{figure*}
\plotone{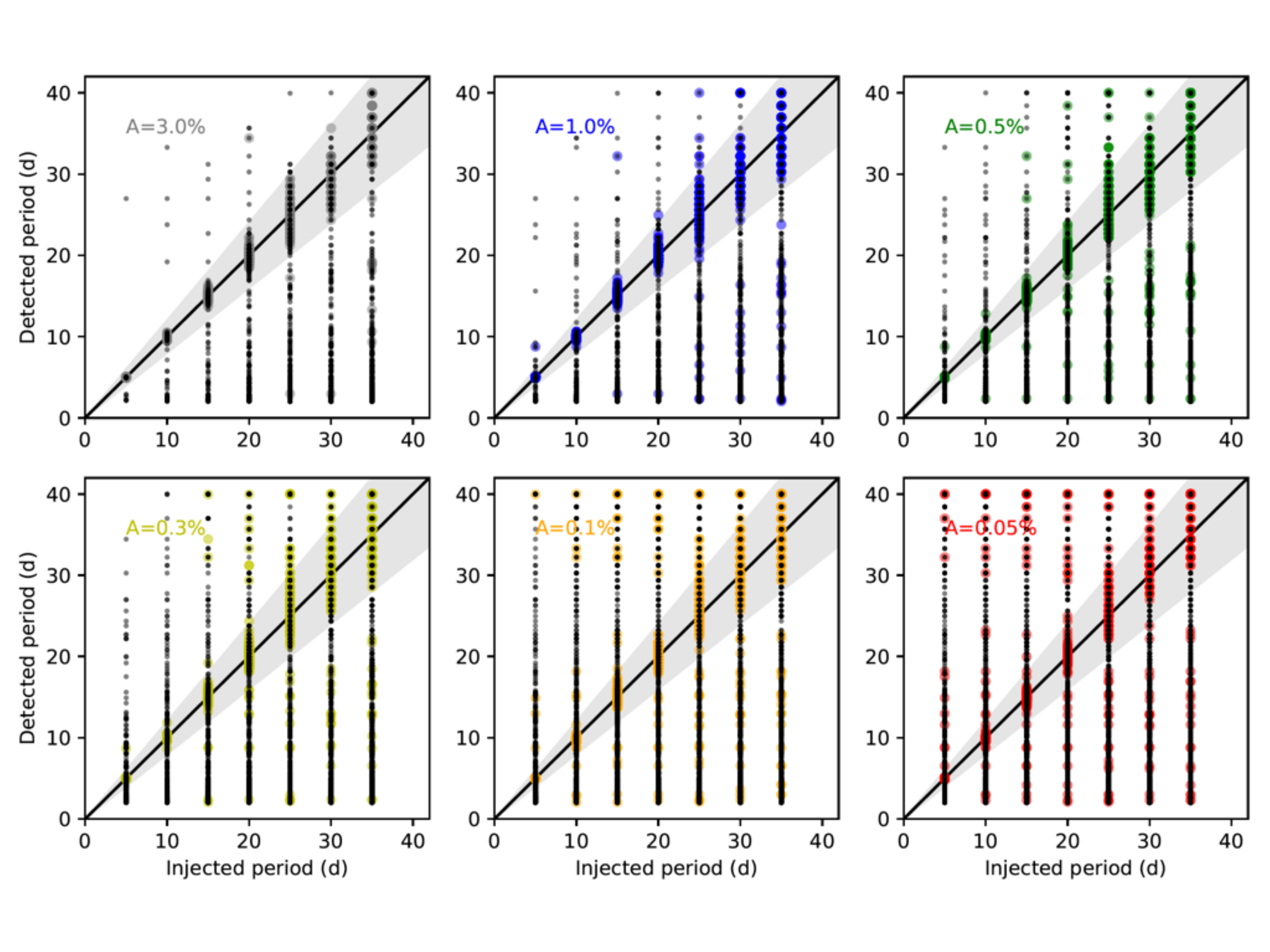}
\caption{Same as Figure~\ref{fig:detvinj1}, but for the SAP sample processed with the CfA pipeline. \label{fig:av_detvinj2}}
\end{figure*}

\begin{figure*}
\plotone{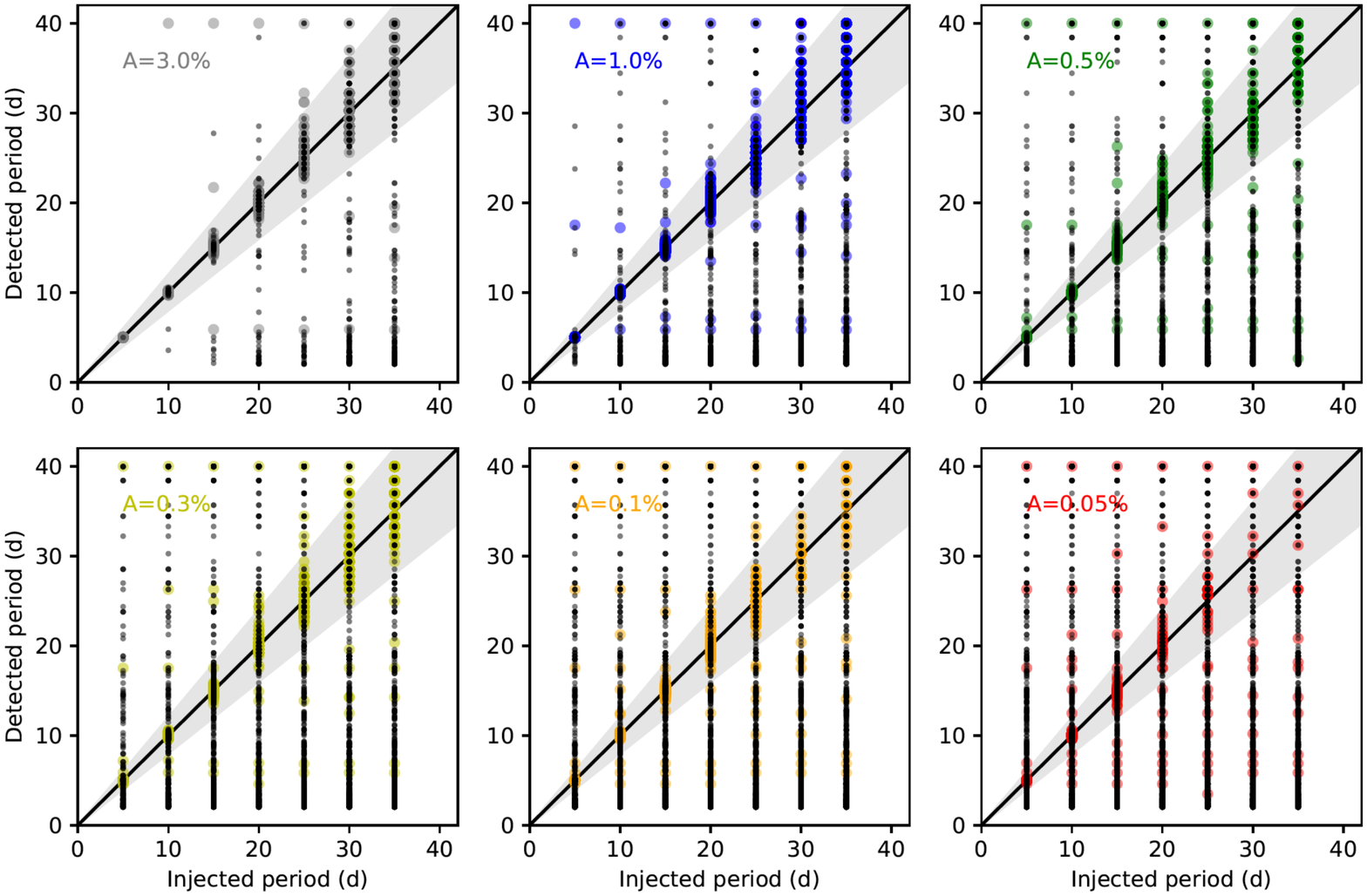}
\caption{Same as Figure~\ref{fig:detvinj1}, but for the SS sample processed with the Oxford pipeline. \label{fig:detvinj3}}
\end{figure*}

\begin{figure*}
\plotone{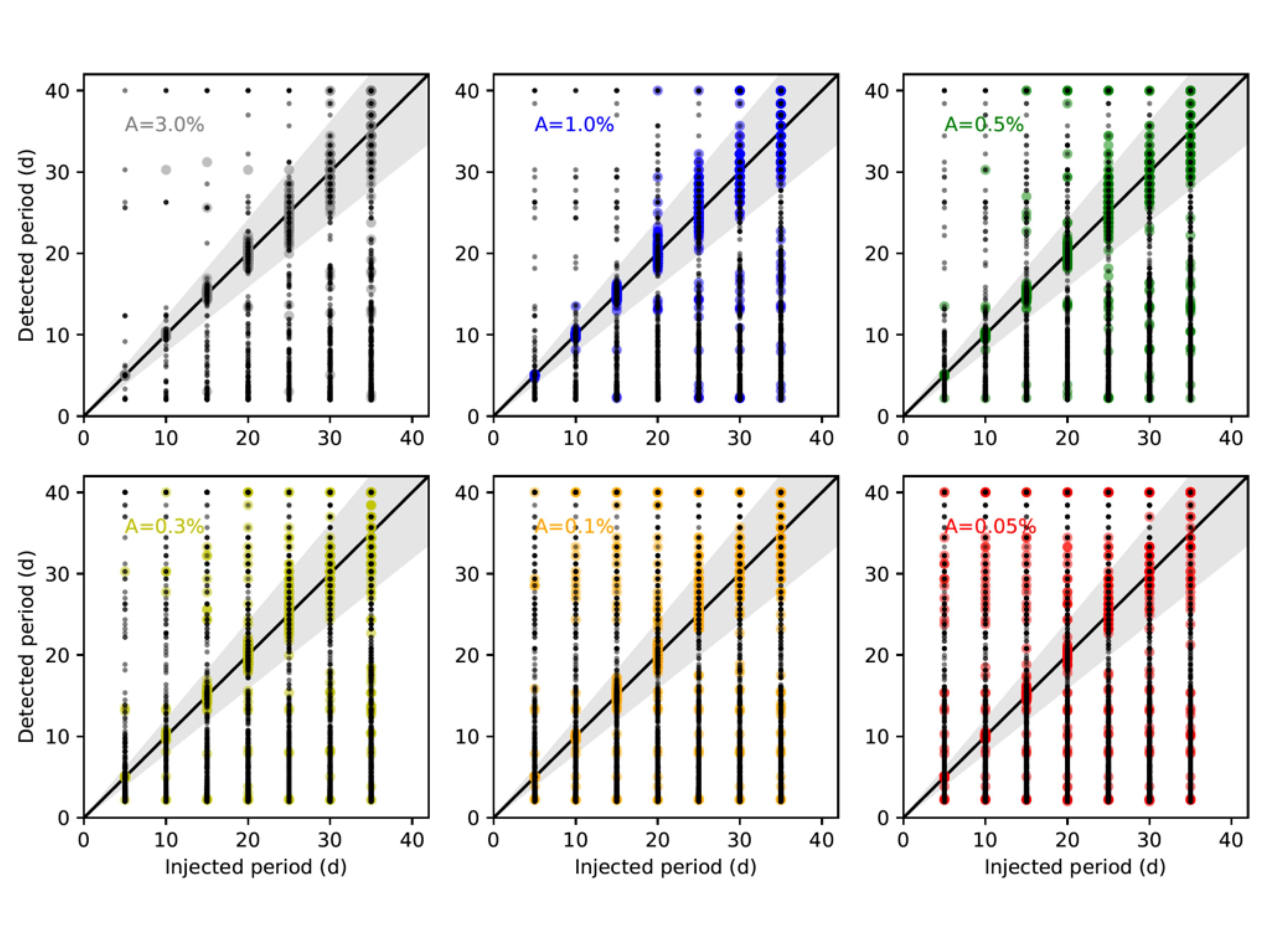}
\caption{Same as Figure~\ref{fig:detvinj1}, but for the SS sample processed with the CfA pipeline. \label{fig:av_detvinj3}}
\end{figure*}

Figures~\ref{fig:detvinj1} through \ref{fig:av_detvinj3} For the hot star, SAP, and SS samples from the Oxford and CfA pipelines.  Each of these figures is made up of 6 panels, corresponding to the injected amplitudes of 3.0, 1.0, and 0.50\% (top row), and 0.3, 0.1, and 0.05\% (bottom row).  Within each panel, the colored circles represent detections, while the cases that did not pass the detection threshold are shown as smaller black dots. The gray shaded area in each plot shows the period validity range, though recall that a $\pm 20\%$ phase match is also required. Phase considerations aside, a black point in the shaded area is a missed valid detection, whereas a colored circle outside the shaded area is an invalid detection.  The stars marked as intrinsically variable were excluded from these figures.

\subsection{Completeness and Reliability} \label{subsec:comprel}

\begin{figure*}
\includegraphics[width=1\linewidth]{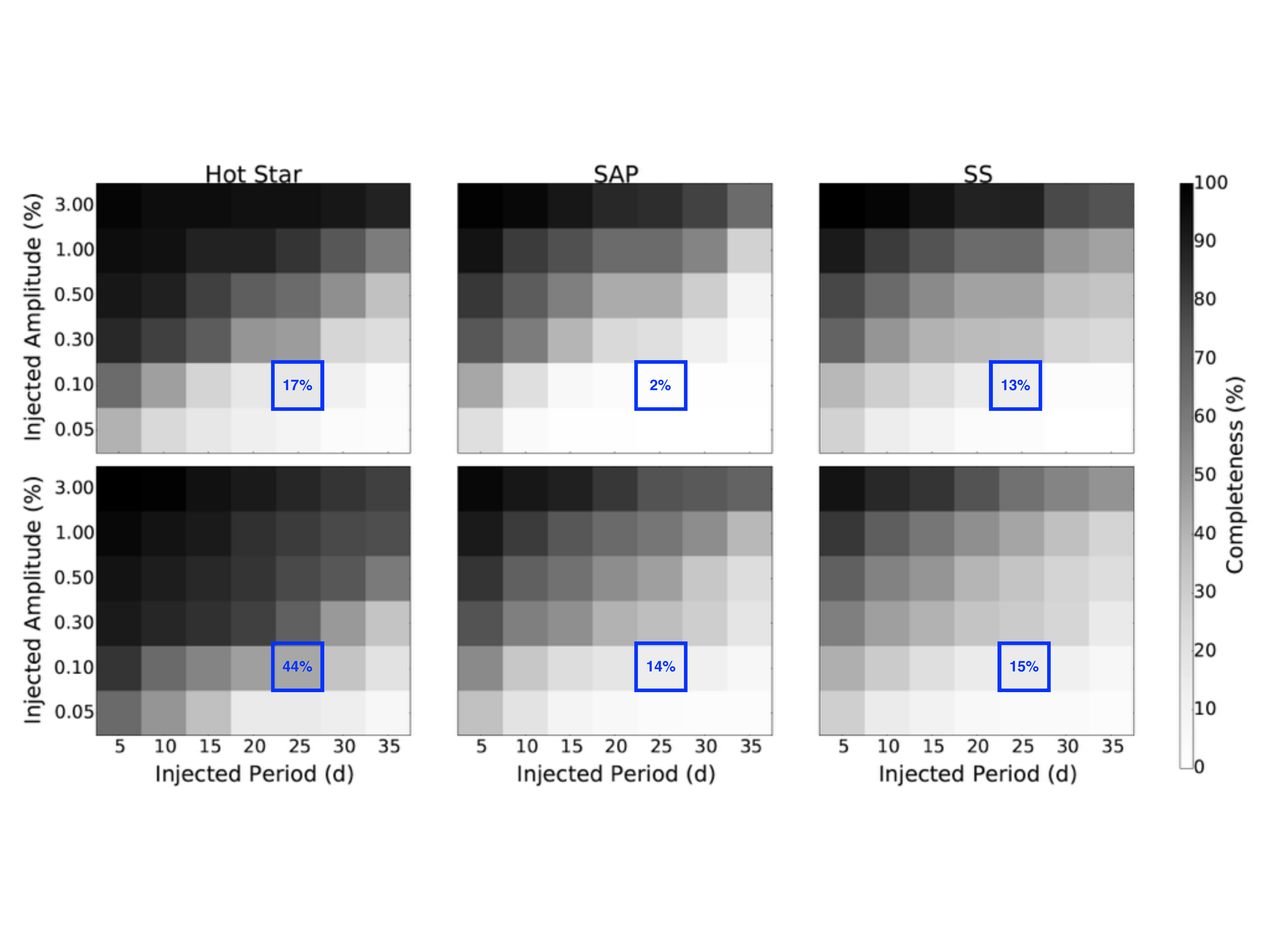} \caption{Completeness tables for the hot star (left), SAP (centre), and SS samples (right), detrended with the Oxford pipeline (top row) and the CfA pipeline (bottom row). The grids have the injected period bin along the x-axis and the injected amplitude along the y-axis. The completeness for the solar case is highlighted in blue for each dataset. \label{fig:completeness}}
\end{figure*}



The completeness results for the Oxford and CfA pipelines are shown in the top and bottom rows of Figure~\ref{fig:completeness}, respectively.  We have identified in blue the completeness for the solar case---where the injected amplitude is 0.1\% and the injected period is 25\,d---for each dataset.  Tables~\ref{tab:hotcomplete} through \ref{tab:ss_av_complete} in Appendix \ref{sec:result_tables} record the completeness values and uncertainties for each injected period-amplitude bin for all test samples from each pipeline.

\begin{figure*}
\includegraphics[width=1\linewidth]{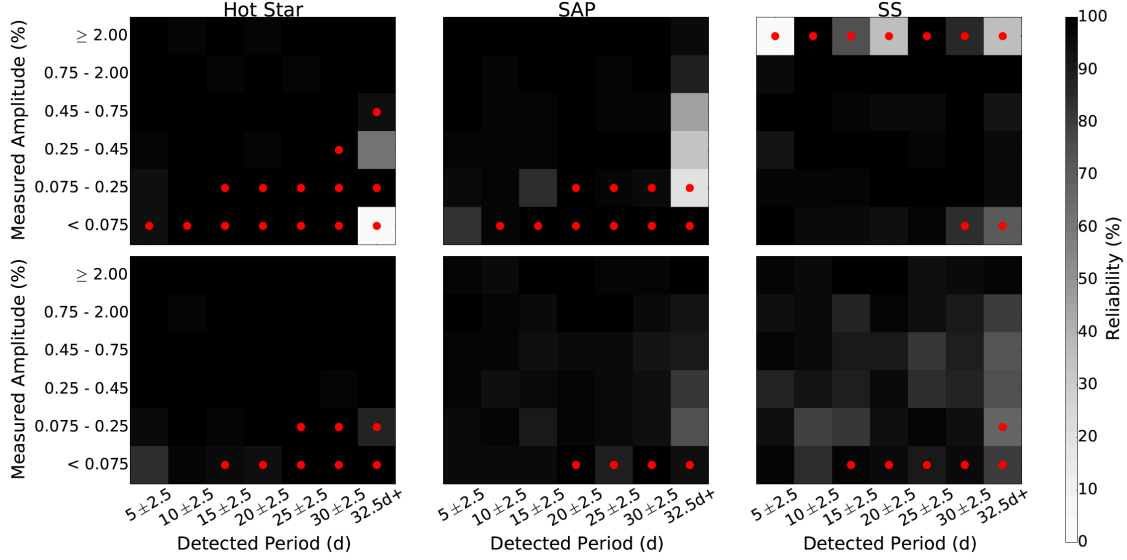} \caption{Reliability for the hot star (left), SAP (centre), and SS samples (right), detrended with the Oxford pipeline (top row) and the CfA pipeline (bottom row). While the bins shown here roughly match those in Figure~\ref{fig:completeness}, here they represent \emph{detected} rather than injected period and amplitude. While the same number of injections was carried out for every injected period and amplitude, the measured and injected values do not necessarily match, so the number of detections in each bin varies.  The red circles represent where there were less than 30 detections in the hot star sample and less than 100 in the SAP and SS samples. \label{fig:reliability}}
\end{figure*}

The reliability results for the Oxford and CfA pipelines are shown in the top and bottom rows of Figure~\ref{fig:reliability}, respectively.  The bins with red dots are those with relatively few detections.  We consider bins with 100 detections or less in the SAP and SS samples and with 30 or less in the hot star sample as insignificant.  All three samples, from both pipelines, have very few detections in the longest period, lowest amplitude bins. The corresponding (generally very high) reliability values therefore have large uncertainties and should not be treated as very meaningful.  The actual values with uncertainties are presented in Tables~\ref{tab:hotreliability1} through \ref{tab:ss_av_reliability1} in Appendix \ref{sec:result_tables}.  

\begin{figure*}
\includegraphics[width=1\linewidth]{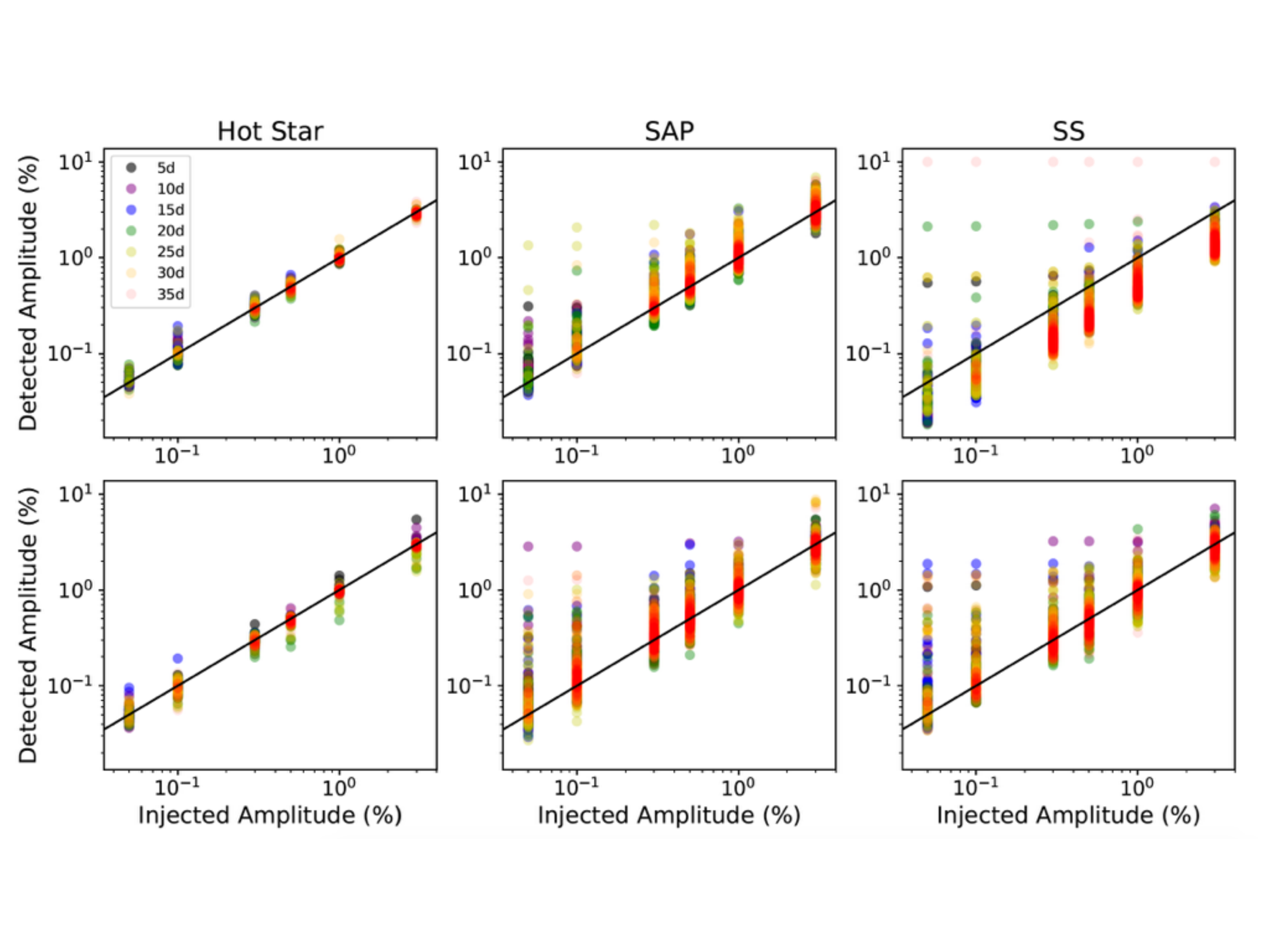} 
\caption{Injected vs. detected amplitude for the valid detections in the hot star (left), SAP (centre), and SS samples (right), detrended with the Oxford pipeline (top row) and CfA pipeline (bottom row). The associated injected periods are marked in different colors.  \label{fig:ox_ampvamp}}
\end{figure*}


Finally, Figure~\ref{fig:ox_ampvamp} shows the injected versus detected amplitudes for valid detections from the Oxford (top) and CfA (bottom) pipelines.  The detected amplitudes can differ significantly from the injected ones for several reasons.  First, the light curves into which the signals were injected may already contain some power at the corresponding period.  Second, the intrinsic noise levels of the light curve will affect the amplitude of the flux as a whole.  Lastly, the systematics correction steps can alter the injected signal, in some cases leading to measured amplitudes that are smaller than the injected ones (e.g. the Oxford SS). 

\section{Discussion} \label{sec:discussion}
We now discuss the results from our injection tests, starting with a brief evaluation of the effects of periodogram normalization in light of the completeness and reliability results in Section \ref{subsec:normal}.  We then look at the `ideal case' of the hot star results in Section \ref{subsec:hot_discuss} before looking at the SAP and SS samples in Section \ref{subsec:sap_ss_discuss}.  Next, we provide an overall comparison of the samples and the pipelines in Section \ref{subsec:compare_stuff}.  Finally, we will discuss the implications of the injection tests on M67 rotation studies using K2 data.

\subsection{Residual long-term trends}
\label{subsec:normal}

The median periodograms shown in Figure~\ref{fig:normal} in Section~\ref{subsubsec:per_norm} tell us a lot about the residual trends present in the light curves after full processing from both the Oxford and CfA pipelines.  For the SAP and SS samples, the median periodogram rises steeply for periods above 20\,d, peaking around $\sim50$ -- $55$\,d and $\sim40$ -- $45$\,d, respectively, before decreasing again and flattening off for periods $>80$\,d (to which we expect little or no sensitivity in K2 light curves anyhow).  While evident in both pipelines, this feature is much more dramatic in the Oxford pipeline.  Importantly, the power present at periods relevant to a study of M67 in both hot star samples shows that this is largely non-astrophysical, despite our steps for systematic correction.  Furthermore, the differences between the SAP and SS samples are significant: our \emph{a priori} expectation was that the SS sample might be more problematic than the SAP one due to increased crowding in the densest parts of M67.  However, the median periodograms tell a different story, with considerably more residual power in the 30 -- 70\,d period range in the SAP than the SS light curves for both pipelines.

If the raw periodograms were used for period detection, these residual trends would lead to numerous false detections at mid-to-long periods (see Appendix~\ref{subsec:nonorm}). Our normalization procedure designed to avoid this seems successful, since we record consistently high reliability wherever we have a significant number of detections (see Figure~\ref{fig:reliability} and Tables~\ref{tab:hotreliability1} to \ref{tab:ss_av_reliability1} in Appendix~\ref{sec:result_tables}). However, normalization also suppresses the detection of real signals at longer periods, as is apparent in Figure~\ref{fig:completeness}.  While we may avoid the trap of lingering systematics by normalizing the periodograms, this comes at the potential cost of real astrophysical signals.

\subsection{The Hot Star Sample}
\label{subsec:hot_discuss}

The set of 89 hot stars is an ideal test case, as they should not show significant rotational modulation beyond about 5\,d.  Therefore, we can be confident that if we see a long period where we did not inject one, it is most likely the result of lingering systematics, to which we alluded above.  For the Oxford pipeline, we were left with 66 stars after removing stars found to have instrinsic variability (i.e. they likely have an astrophysical signal that could interfere with the injection test results) using our detection threshold on the non-injected versions of the light curves.  Of the 23 stars marked as variable, 19 had detected periods less than 10\,d and the rest had periods less than $\sim$16\,d, apart from one (see Figure~\ref{fig:hotstarhist}).  In the CfA pipeline, we used 70 stars in the analysis.  All 19 variables had periods under 12\,d.  These numbers further enforce the fact that our normalization works reasonably well for this particular dataset in both pipelines, though the presence of flagged variables with periods longer than 5\,d shows that the set of hot stars may also be contaminated by background stars, or that there are a number of cooler F stars in the sample, but it is good enough for our purposes.

As we can see in Figures~\ref{fig:detvinj1} and \ref{fig:av_detvinj1}, we do a good job at recovering the injected periods from both pipelines in the hot star sample.  Some of the invalid detections likely come from variables --- possibly pulsators --- not removed from the sample but with low amplitudes. Where these start to become detections at the lowest injected amplitudes, the injected signal may be `boosting' (i.e. increasing the amplitude of the intrinsic signal) them just enough to be detected.  Even for very regularly sampled data like K2, injecting a signal at a given period affects the periodogram at other frequencies in a complex way.  However, a number of the invalid detections could also be the result of the normalization, which promotes shorter periods at the expense of longer ones:  when dividing out the median periodograms, the much lower power seen at short periods increases short-period significance relative to longer periods.

The hot star completeness values for both pipelines can be seen in the leftmost panels of Figure~\ref{fig:completeness} and in Tables~\ref{tab:hotcomplete} and \ref{tab:hot_cfa_complete} in Appendix~\ref{sec:result_tables}.  As expected from simple signal-to-noise arguments, we see completeness decrease with increasing injected period and decreasing injected amplitude for both pipelines.  We point out that the best-case scenario for recovering a solar-like signal of 25\,d and 0.1\% amplitude is $\sim$45\% from the hot star dataset.  However, the limited size of the hot star sample means that the individual completeness values are somewhat uncertain ($\pm$ $\sim$12\%) for both pipelines.

The reliability in the hot star sample from either pipeline is consistently high ($>$90\%).  High reliability means that we can typically trust a detection made within a measured amplitude and period range.  We can see from Figure~\ref{fig:reliability} that where we have a significant number of detections (more than 30 for the hot star samples), we can be fairly confident in our results, though again, the uncertainties are relatively large for the hot star samples.  The high reliability also indicates that normalization did not introduce low frequency signals in light curves that did not have a strong intrinsic signal in the first place, again reinforcing the validity of this step.

\subsection{SAP and SS Completeness and Reliability}
\label{subsec:sap_ss_discuss}

We have shown that the hot star sample is important for validating our injection test procedures and highlighting the presence of non-astrophysical power around 25\,d and longer in the K2 Campaign 5 light curves.  We now discuss the results from the SAP and SS samples, which specifically illustrate the complexity of a period search in M67.  Figures~\ref{fig:detvinj2} to \ref{fig:av_detvinj3}, which show the detected versus injected periods for the SAP and SS datasets from both pipelines, are a lot messier than their hot star counterparts, most likely due to the larger sample sizes and more diverse stellar populations within.  However, like the hot stars, detections become more difficult and less reliable as the injected period increases and the injected amplitude decreases.  In addition, lingering power around 10 -- 20\,d seems to exist in the SAP and SS light curves.  This power was either not originally strong enough to mark the light curves as variables until boosted by a lower-amplitude, injected signal, or it could be the result of half-period harmonic measurements from the Lomb-Scargle periodogram.

The completeness results for both the SAP and SS samples are in the middle and far right panels, respectively, in Figure~\ref{fig:completeness}.  As with the hot stars, we see the same general trend of diminishing completeness with increasing injected period and decreasing amplitude, but it has been exacerbated.  For both samples from either pipeline, collapsing the figures onto either axis, the completeness falls around or below 50\% for amplitudes $\leq$0.50\% and periods $\geq$20\,d, showing just how hard it is to detect long-period, low-amplitude signals in either case.  Of particular importance is the solar case, highlighted in blue in Figure~\ref{fig:completeness}, where the injected period is 25\,d and the amplitude is 0.10\%.  Critically, the completeness here is $\sim$15\% or lower for both the SAP and SS samples in either pipeline.  Even with the best-case scenario of perfect sinusoidal signals, solar variability, which we expect in M67, is clearly difficult to find with our detection criteria.

The middle and far right panels of Figure~\ref{fig:reliability} present the reliability for the SAP and SS samples.  As with the hot stars, where there is a significant number of detections ($>100$ for the larger datasets), the reliability for the SAP and SS samples rarely drops below 90\% for both pipelines.  This is encouraging, as it shows that the procedure we have developed should lead to relatively few false alarms, i.e. where we do get a detection, we can generally trust the period measurement.

There are a few striking features in Figure~\ref{fig:reliability}, however.  Detecting long periods (particularly at $\sim$35\,d) in the Oxford SAP sample appears to be not only more difficult than the CfA pipeline, but also less reliable. The low reliability in this period range could either be a result of periodogram normalization or a failure to correct the residual long-term trends in the data, which the CfA pipeline does better.  The Oxford SAP sample also has lower reliability in the shortest period, lowest amplitude bin.  This could be an effect of promoting short-period signals when dividing out the median periodogram.  We experimented with adding a `floor' to the Oxford median periodograms, by which we set any value below a given threshold to that floor in an attempt to reduce this effect.  We tested floor values of 0.005 and 0.01.  While these did improve reliability in the short period, low amplitude bin, the floors tended to reduce reliability overall in all three Oxford samples, particularly the hot star and SS sets, and especially around mid-range periods.  Thus, we decided to exclude a floor from the study to avoid further artificial effects.

One surprising feature in the SS samples is the lack of detections across all periods at amplitudes of 2.0\% and greater for the Oxford pipeline, even though many 3.0\% signals were injected. This can be seen in both Figures~\ref{fig:reliability} and \ref{fig:ox_ampvamp}.  If we then compare the number of SS detections from one pipeline to another in the amplitude ranges 0.75\% -- 2.00\% and 0.075\% -- 0.25\% in Tables~\ref{tab:ssreliability1} and \ref{tab:ss_av_reliability1} in Appendix~\ref{sec:result_tables}, there are systematically more detections in the Oxford pipeline for these ranges than in the CfA pipeline.  This indicates that the amplitudes of the Oxford SS sample are suppressed, primarily after the PCA.  Though the PCA may also suppress amplitudes in the Oxford SAP sample, it is more evident in the SS sample due to the high noise amplitude of the SS PCs compared to the SAP PCs.  The amplitude suppression could also indicate that the Oxford pipeline removes some intrinsic variability in addition to the systematics.  However, it seems that despite the suppression, the signals remain intact enough to be detected at a similar rate, just at lower amplitudes.

\subsection{Comparing the Samples and Pipelines} \label{subsec:compare_stuff}
In both pipelines, the hot star sample has the highest overall completeness and reliability.  This is unsurprising given its smaller sample size and lack of diversity in terms of variability.  Due to the general absence of rotational modulation and other competing signals in the original light curves, the hot stars should also better preserve the injected sinusoids.  Finally, the average brightness of the hot star dataset is $K_{p}=11.0$, while the combined average of the SAP and SS datasets is $K_{p}=15.3$, and the number of valid detections generally decreases with increasing magnitude. 

The Oxford SS completeness is slightly lower than the Oxford SAP at short periods and large amplitudes, but it is better for periods $\geq15$\,d and amplitudes below about 0.50\%.  In addition, the SS sample is more reliable for both the longest and shortest detected periods.  These differences are likely due to the separate light curve extraction methods in the Oxford pipeline.  The SAP sample is extracted with pixellated apertures, which may not be very optimal for a crowded field, even in the outer regions of the cluster.  The SS sample, however, uses deblended, circular apertures, making it better able to deal with crowding.

On the other hand, the CfA SAP completeness and reliability are very comparable (or slightly higher) across most bins when compared against the CfA SS sample.  The slight advantage to the SAP sample is probably due to less crowding, as the two datasets are extracted and processed the same way in the CfA pipeline.  In addition, the larger size of the SAP sample means that there is a bigger matrix from which to characterize and correct the common-mode trends via PDC-MAP (or PCA, in the case of the Oxford pipeline).  This could partially help explain the generally higher completeness values from the SAP sample in either pipeline, as well as the higher reliability in the case of the CfA pipeline.

\begin{figure*}
\includegraphics[width=1\linewidth]{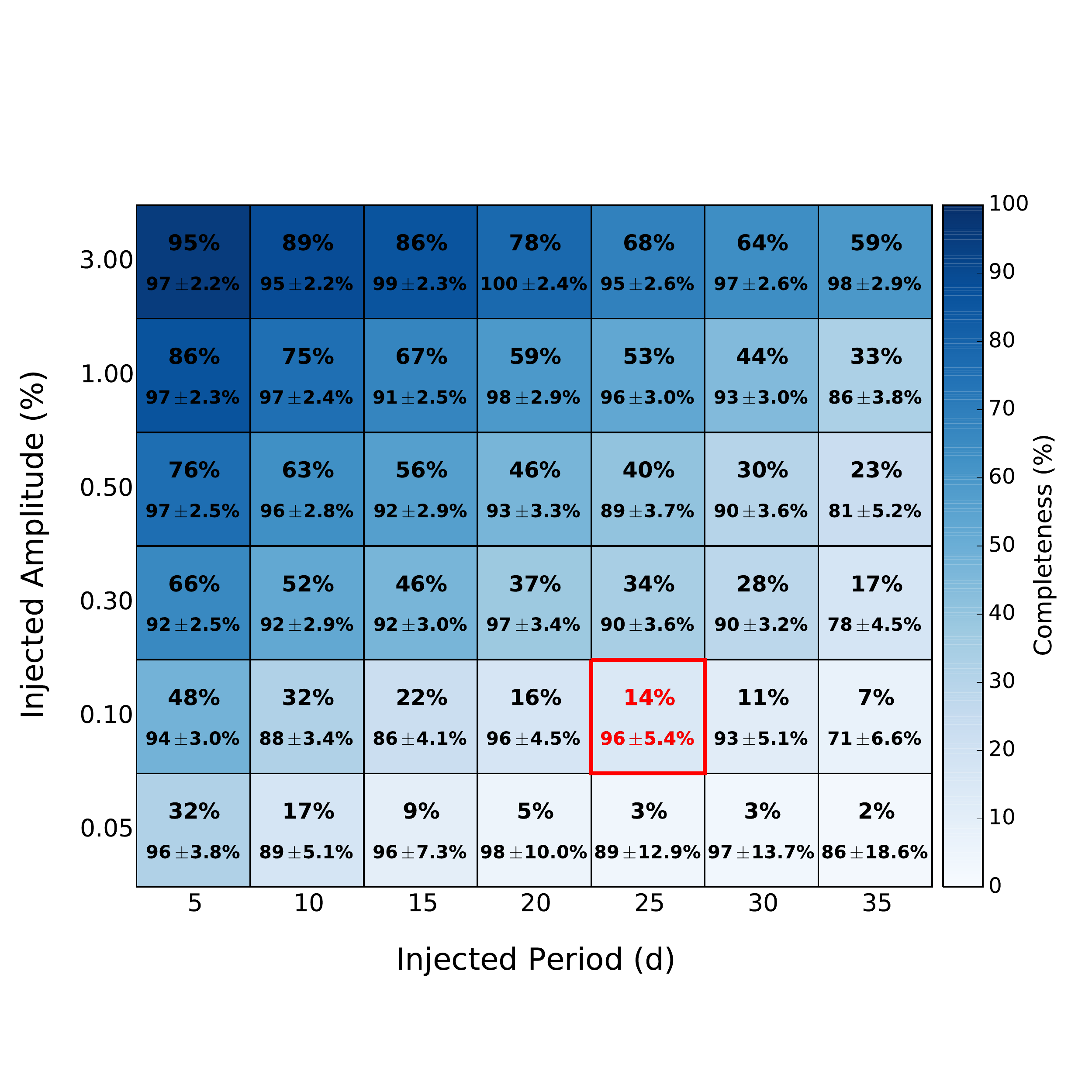} \caption{Average completeness (top number) and the roughly associated reliability (bottom number with uncertainty) for the CfA SAP and SS samples, with the solar case highlighted in red. The uncertainty for the completeness values is $\pm$2.1\%.  Note that the reliability values do not perfectly match the injected amplitudes and periods but are representative of the ranges shown in Figure~\ref{fig:reliability}.  \label{fig:summary}}
\end{figure*}

With respect to completeness, the CfA pipeline seems to perform slightly better than the Oxford pipeline, particularly in the hot star and SAP samples.  Thus, Figure~\ref{fig:summary} provides the average completeness (and associated reliability in smaller text) from the CfA SAP and SS samples as a summary of the best we can do with the K2 Campaign 5 M67 data, with the solar case highlighted in red.  The differences between the Oxford and CfA pipelines are rooted in the approaches each takes to produce corrected light curves from raw K2 data.  The moving apertures and deblending procedures during light curve preparation give the Oxford pipeline an edge in certain cases, particularly with the SS sample, but the larger, more optimized aperture selection from the CfA pipeline is superior elsewhere.  While the Oxford pipeline typically better removes systematics from strongly variable stars, most M67 targets won't be sufficiently variable on short timescales, so the advantage is minimal.  Finally, the more sophisticated PDC-MAP outperforms the relatively crude PCA of the Oxford pipeline in the common-mode systematic removal.  While an ideal pipeline would combine the best elements of the the two, the performance of the existing pipelines is comparable, and both are valid options regarding the analysis of K2 light curves.  Most critically, however, is that for both pipelines, the completeness around the solar case in the SAP and SS samples is at best $\sim$15\%, illustrating that it is very difficult to detect 0.1\% amplitude, 25\,d signals in K2 Campaign 5 M67 data.  

\subsection{Implications for K2 M67 rotation studies} \label{subsec:imply}

\begin{deluxetable*}{lccccccccc}
\tablecaption{Comparison with periods from \citet{2016ApJ...823...16B} \label{tab:barnes_compare}}
\tablecolumns{8}
\tablenum{13}
\tablewidth{0pt}
\tablehead{
\colhead{EPIC} &
\colhead{Barnes Per (d)} &
\colhead{Amp (\%)} & \colhead{Comp (\%)} & \colhead{Rel (\%)} &
\colhead{Ox Per (d)} & \colhead{CfA Per (d)}
}
\startdata
211388204 & 31.8 & 0.32 & 21 $\pm$1.8 & 96 $\pm$3.9 & -- & --  \\
211394185 & 30.4 & 0.78 & 55 $\pm$1.8 & 98 $\pm$2.3 & 15.4 & 15.1 \\
211395620 & 30.7 & 0.47 & 32 $\pm$1.8  & 95 $\pm$3.1 & -- & -- \\
211397319 & 25.1 & 0.31 & 29 $\pm$1.8  & 98 $\pm$3.2 & -- & -- \\
211397512 & 34.5 & 0.73 & 34 $\pm$1.8 & 68 $\pm$4.1 & 16.4 & -- \\
211398025 & 28.8 & 0.29 & 21 $\pm$1.8 & 96 $\pm$3.9 & -- & -- \\
211398541 & 30.3 & 0.51 & 32 $\pm$1.8 & 95 $\pm$3.1 & -- & -- \\
211399458 & 30.2 & 0.66 & 32 $\pm$1.8 & 95 $\pm$3.1 & -- & -- \\
211399819 & 28.4 & 0.31 & 21 $\pm$1.8 & 96 $\pm$3.9 & -- & -- \\
211400500 & 26.9 & 0.30 & 29 $\pm$1.8 & 98 $\pm$3.2 & -- & -- \\
211406596 & 26.9 & 0.36 & 29 $\pm$1.8 & 98 $\pm$3.2 & -- & -- \\
211410757 & 18.9 & 0.14 & 11 $\pm$1.8 & 99 $\pm$5.8 & -- & -- \\
211411477 & 31.2 & 0.20 & 21 $\pm$1.8 & 95 $\pm$11.9 & -- & -- \\
211411621 & 30.5 & 0.30 & 21 $\pm$1.8 & 96 $\pm$3.9 & 15.8 & 16.1 \\
211413212 & 24.4 & 0.22 & 29 $\pm$1.8 & 97 $\pm$8.0 & -- & -- \\
211413961 & 31.4 & 0.26 & 21 $\pm$1.8 & 96 $\pm$3.9 & -- & -- \\
211414799 & 18.1 & 0.17 & 11 $\pm$1.8 & 99 $\pm$5.8 & -- & 9.0 \\
211423010 & 24.9 & 0.29 & 29 $\pm$1.8 & 98 $\pm$3.2 & -- & 22.2 \\
211428580 & 26.9 & 0.33 & 29 $\pm$1.8 & 98 $\pm$3.2 & -- & -- \\
211430274 & 31.1 & 0.40 & 32 $\pm$1.8 & 96 $\pm$3.9 & -- & -- \\
\enddata
\tablecomments{The columns from left to right are EPIC, the associated period reported from \citet{2016ApJ...823...16B}, our estimate of the peak-to-peak amplitude for the light curve, the estimated completeness and reliability statistics based off the Barnes period and amplitude, and the Oxford and CfA periods where we have detections using $C=4$.}
\end{deluxetable*}

\begin{deluxetable*}{lccccccccc}
\tablecaption{Comparison with detected periods from \citet{2016MNRAS.463.3513G} \label{tab:gonz_compare}}
\tablecolumns{8}
\tablenum{14}
\tablewidth{0pt}
\tablehead{
\colhead{EPIC} &
\colhead{Gonzalez Per (d)} & \colhead{Amp (\%)} & \colhead{Comp (\%)} & \colhead{Rel (\%)} &
\colhead{Ox Per (d)} & \colhead{CfA Per (d)}
}
\startdata
211387834 & 15.2 & 1.36 & 74 $\pm$1.8 & 98 $\pm$2.0 & 14.9 & 14.9   \\
211390071 & 25.3 & 1.42 & 63 $\pm$1.8 & 99 $\pm$2.2 & 13.5 & -- \\
211393422 & 28.6 & 1.53 & 55 $\pm$1.8 & 98 $\pm$2.3 & 14.5 & -- \\
211397501 & 12.4, 12.4 & 1.19 & 81 $\pm$1.8 & 98 $\pm$1.9 & 12.2 & 12.5 \\
211397955 & 29.0, 29.2 & 1.36 & 55 $\pm$1.8 & 98 $\pm$2.3 & 14.7 & 14.5 \\
211400662 & 13.5 & 2.04 & 91 $\pm$1.8 & 99 $\pm$1.8 & 13.7 & 13.1 \\
211403852 & 27.3 & 1.23 & 63 $\pm$1.8 & 99 $\pm$2.2 & 13.9 & 13.3 \\
211404310 & 26.3 & 1.17 & 63 $\pm$1.8 & 99 $\pm$2.2 & -- & 13.0 \\
211405671 & 25.6, 28.2 & 2.06, 2.35 & 77, 76 $\pm$1.8 & 98 $\pm$2.0, 99 $\pm$1.9 & 26.3 & -- \\
211407277 & 24.7 & 2.50 & 77 $\pm$1.8 & 98 $\pm$2.0 & -- & 25.0 \\
211408116 & 30.1 & 0.14 & 6 $\pm$1.8 & 95 $\pm$11.9 & 13.5 & -- \\
211408874 & 23.9, 24.5 & 2.12 & 80 $\pm$1.8 & 98 $\pm$2.0 & -- & 23.8 \\
211414974 & 26.8, 29.9 & 1.91, 1.92 & 63 $\pm$1.8 & 97 $\pm$2.1, 98 $\pm$2.3 & 27.0 & -- \\
211424980 & 15.6, 31.2 & 1.26, 1.49 & 74 $\pm$1.8 & 98 $\pm$2.0, 98 $\pm$2.3  & 15.6 & 15.6 \\
211427666 & 28.8 & 1.20 & 55 $\pm$1.8 & 98 $\pm$2.3 & 14.1 & -- \\
211429354 & 25.9 & 0.19 & 8 $\pm$1.8 & 97 $\pm$8.0 & -- & 26.3 \\
211430648 & 32.6 & 0.67 & 16 $\pm$1.8 & 68 $\pm$4.1 & -- & 17.2 \\
211433352 & 13.7, 13.8 & 1.21 & 74 $\pm$1.8 & 98 $\pm$2.0 & 13.9 & 13.5 \\
\enddata
\tablecomments{The columns from left to right are the EPICs from \citet{2016MNRAS.463.3513G} where either of our two pipelines have detections, the associated periods reported from that paper, our estimate of the peak-to-peak amplitude for each light curve, the estimated completeness and reliability statistics based off the Gonzalez period and amplitude, and the Oxford and CfA periods where we have detections using $C=4$. Where there is more than one reported value, the first number comes from the PDCSAP sample from \citet{2016MNRAS.463.3513G}, while the second number is from the K2SC sample.}
\end{deluxetable*}

\begin{figure*}
\includegraphics[width=1\linewidth]{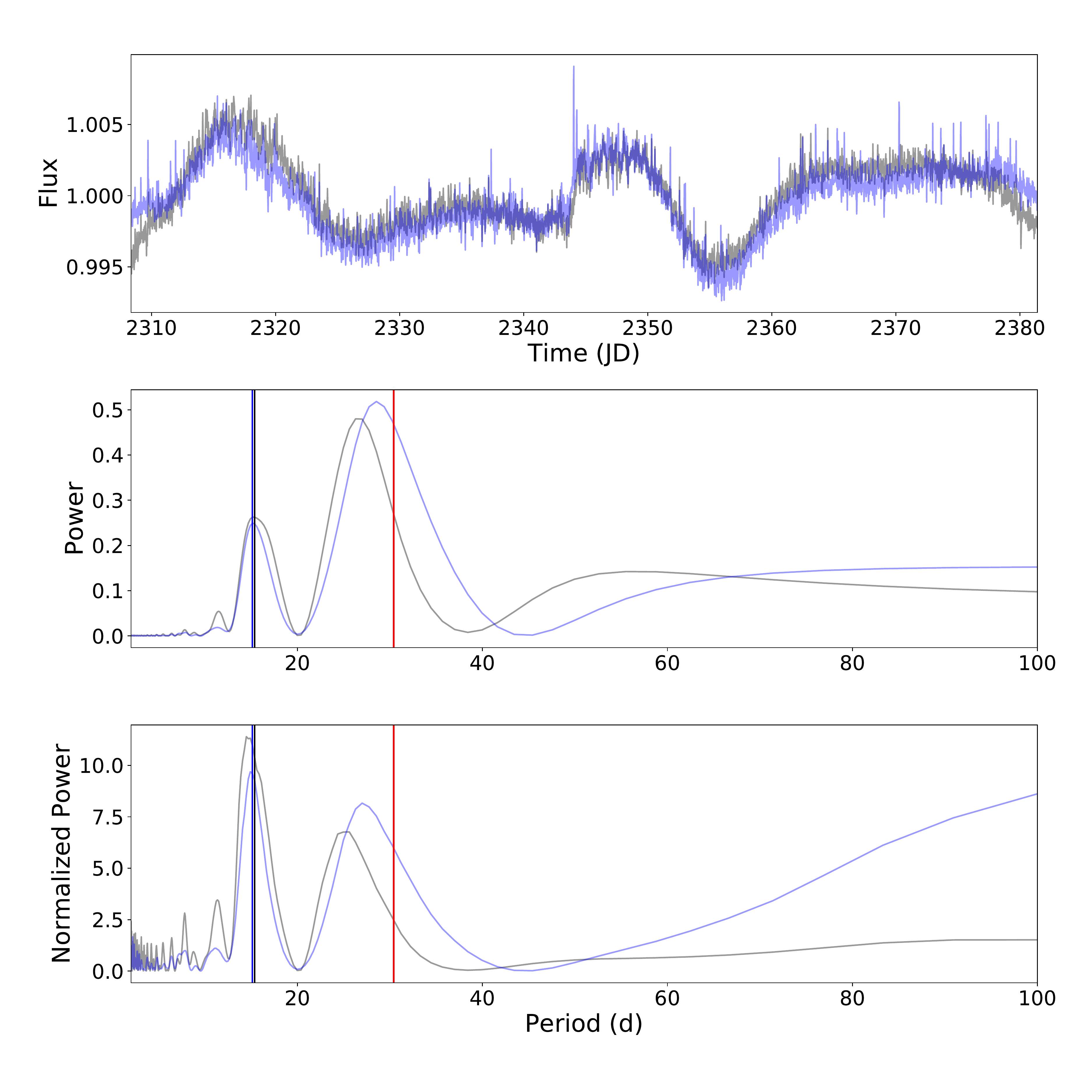} \caption{Light curves and periodograms for EPIC 211394185 from the Oxford (gray) and CfA (blue) pipelines. The top panel displays the light curves, while the middle panel gives the original periodogram power for both pipelines, and the bottom panel shows the normalized periodograms.  In both periodogram panels, the black lines show the final Oxford period, the blue lines show the location of the final CfA period, and the red lines show the location of the period from the Barnes paper. \label{fig:211394185}}
\end{figure*}

\begin{figure*}
\includegraphics[width=1\linewidth]{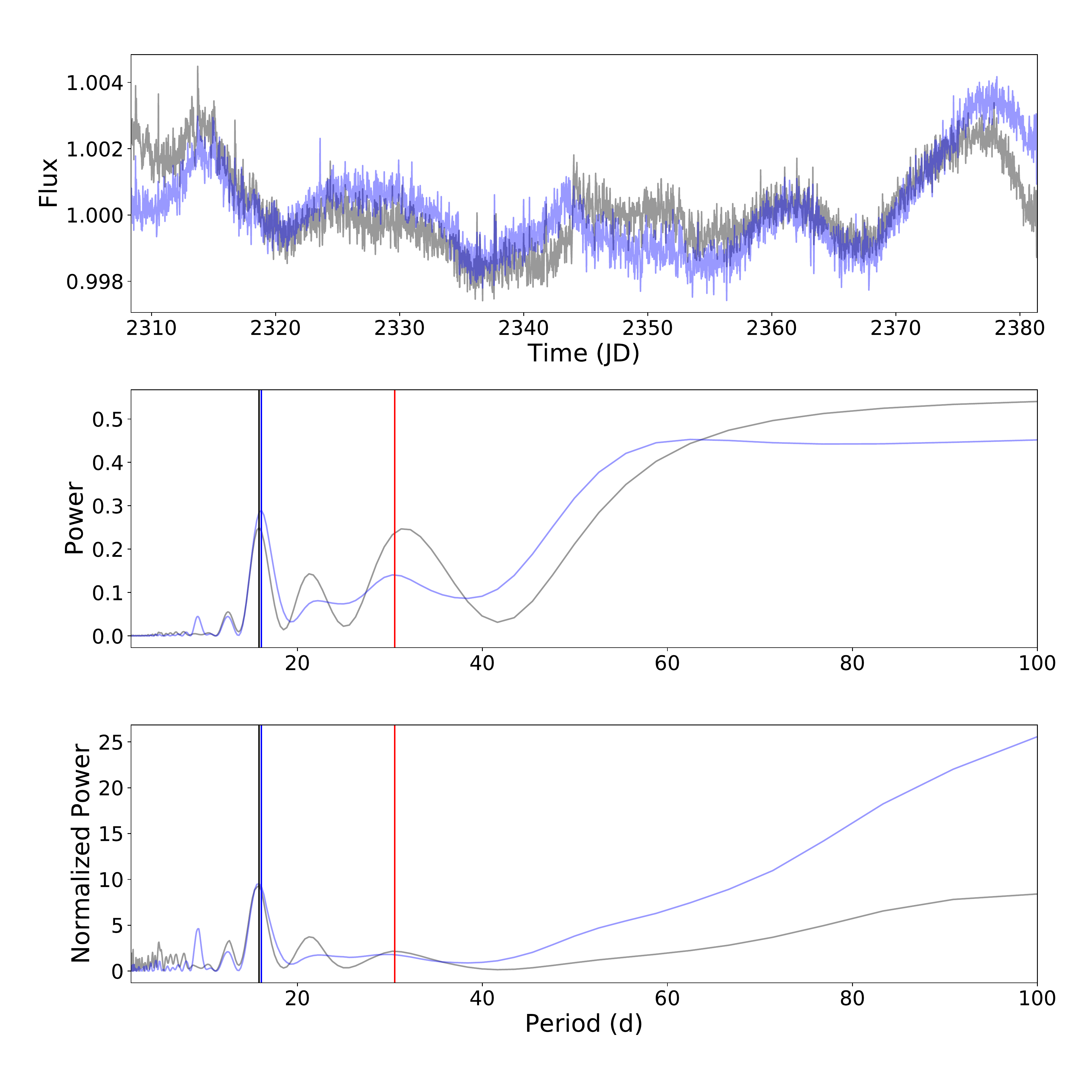} \caption{Same as Figure~\ref{fig:211394185} except for EPIC 211411621. \label{fig:211411621}}
\end{figure*}

The injection test results indicate that, if the Sun is considered typical, measuring rotation periods for solar-like stars in M67 based on their K2 light curves is challenging. Specifically, for the best case scenario of solar-like rotational variability --- 0.1\% amplitude, 25\,d non-evolving sinusoidal signal --- our completeness maxes out at 14\% and 15\% for the SAP and SS samples, respectively.  Considering that detecting non-sinusoidal, evolving spot-modulation will be more difficult, these results suggest that both current and future rotation period measurements in M67, based solely on the K2 Campaign 5 data, must be carefully examined before they are used to guide models of stellar angular momentum evolution and/or calibrate gyrochronology relations. 

We consider the 20 stars with measured rotation periods from \citet{2016ApJ...823...16B} and the stars from \citet{2016MNRAS.463.3513G} where we had detections from either the CfA or Oxford pipeline.  Recall that all of these light curves come from the SAP sample.  Since we do not have amplitude measurements for the versions of the light curves used in either paper, we estimated them by binning the flux of each light curve based on the reported period and taking the median of the difference between the $95^{th}$ and $5^{th}$ percentiles of each bin.  We did this for both the Oxford and CfA versions of each light curve and averaged the values.  We then estimated the completeness and reliability based on the corresponding SAP results from this study.  We averaged the Oxford and CfA completeness for the injected period and amplitude bins closest to the determined values.  If the calculated amplitude was halfway between the injected amplitudes, we used the higher amplitude bin.  We also used the average Oxford and CfA reliability from the appropriately ranged bins into which the reported periods and estimated amplitudes fell.  

The results are shown in Tables~\ref{tab:barnes_compare} and \ref{tab:gonz_compare}.  The completeness for the reported periods from \citet{2016ApJ...823...16B} is low, averaging at about 27\%, with only one period falling into a range above 50\% (EPIC 211394185).  The corresponding reliability is high, only dipping below 95\% in one instance (EPIC 211397512).  However, these values are a bit misleading, as they only really apply to those cases where we would actually get a detection according to our criteria, i.e. using a threshold factor of $C=4$.  If we have a detection in an area with low completeness, then the result is generally trustworthy.  Therefore, in Table~\ref{tab:barnes_compare}, we also show the period from the normalized Lomb-Scargle for the corresponding, non-injected light curves from the Oxford and CfA pipelines where we had a detection based on our threshold (i.e. where the star was classified as a `variable').  For this sample of stars, the Oxford pipeline only has 3 detections, while the CfA has 4.  Almost all of these detections appear to be rough harmonics of the published result from \citet{2016ApJ...823...16B}, while the CfA period for EPIC 211423010 is close to the Barnes value.  

Both the Oxford and CfA pipelines have detections for EPICs 211394185 and 211411621 from the Barnes sample.  Figures~\ref{fig:211394185} and \ref{fig:211411621} show the Oxford and CfA versions of these two light curves, respectively, along with the original periodograms in the middle panels and the normalized versions in the bottom panels. While obviously variable, a clear period is difficult to find by eye in either case, though it appears that the normalization suppressed any power at a period that would be comparable to the Barnes result.  The normalization step is important for avoiding false positives, however, as we have illustrated with the presence of long-term trends in the median periodogram of the hot star samples.  We recognize that \citet{2016ApJ...823...16B} used several different period detection methods other than the Lomb-Scargle periodogram.  However, the concern here is obviously the underlying signal, and the lack of detections from the Oxford and CfA pipelines, as well as the lack of agreement with the Barnes result where we do have one, show just how challenging it is to measure solar-like signals in the K2 Campaign 5 data.

Looking at Table~\ref{tab:gonz_compare}, we have a list of 18 EPICs from either the `PDCSAP' or `K2SC' samples given in \citet{2016MNRAS.463.3513G} where we had a detection from at least one of our pipelines.  The average completeness is $\sim$60\%, which is higher than the Barnes sample, and the reliability is $\geq$95\% (except in one case, EPIC 211430648), but the stars listed here are only a small fraction of those published by \citet{2016MNRAS.463.3513G}.  The PDCSAP sample in total has 98 reported periods; we detected 16 from this sample.  The K2SC sample had 40 periods; we detected 9.  In addition, notice that none of the stars in Table~\ref{tab:barnes_compare} show up in \ref{tab:gonz_compare}, meaning that of the 9 stars reported by \citet{2016MNRAS.463.3513G} that overlap with \citet{2016ApJ...823...16B}, neither the Oxford nor the CfA pipeline records a detection.  However, as with the Barnes detections, the CfA and Oxford results agree, and they are either in agreement with the Gonzalez value, or they are rough harmonics.  The harmonics again highlight ambiguity in the light curves, while the general lack of detections from the Gonzalez sample --- especially where reported from Barnes --- reinforces the conclusions we have already made about finding solar signals in M67 with K2 data.

We can use the completeness and reliability statistics from this work to guide future re-evaluations of M67 periods from K2 data.  Reliability helps establish amplitude and period thresholds for detecting true signals in M67.  We can probably trust detections in a measured period and amplitude bin with reliability of 80\% or higher, suggesting that 1 or fewer out of every 5 measurements is likely to be unreliable.  However, we need to take into account completeness, as we simply cannot say anything about the true number of stars with periods and amplitudes corresponding to a given bin without it, since completeness tells us what fraction of potential detections we miss.  For example, if we have three detections in a bin where the completeness was 50\%, we know that there were approximately 3 other stars in that range that we missed.

It is important to recall that these sinusoidal injection tests represent a `best case' scenario: real rotation signals are non-sinusoidal and evolve over time, and both of these factors are likely to affect their detectability in a negative way. Therefore, although it is of course possible that better light curve extraction, detrending, and period search methods might be devised than those we have used here, it seems rather unlikely that the detectability of true rotation signals will improve much beyond what we have shown. In short, the K2 Campaign 5 data may allow the measurement of some rotation periods for main sequence stars in M67, but only if they are relatively active and display amplitudes somewhat larger than the active Sun. 

\section{Conclusions and Future Work} \label{sec:conclusion}
The open cluster M67, recently observed in Campaign 5 of the NASA K2 mission, offers the unique chance to measure rotation periods for solar-age stars.  This means that we have the opportunity to fill in a much-needed gap in the calibration of the age-mass-rotation period relationship which forms the basis for the field of gyrochronology and serves as an age-dating method for main sequence stars.  However, our physical understanding behind the empirical observations that have driven gyrochronology is still evolving, and certain studies suggest inconsistencies between the ages that we would predict via gyrochronology and the results of other methods, such as asteroseismology, for older stars.  M67, therefore, provides a testing ground both for gyrochronology and new theories exploring the possible weakening of the magnetic fields that induce angular momentum loss in stars.  Because of the immense scientific potential of M67, we want to ensure that we have a full understanding of the limits of the data we are using to measure the rotation periods.  Though K2 does have relatively high precision, especially compared to ground-based observations, the data still suffers from stubborn systematic features, and the $\sim$75\,d observation window means that it will be challenging to identify periods of 25\,d or longer, which we are expecting for a cluster the approximate age of the Sun.  In addition, the crowded field will make the task even more difficult.  It is the goal of this study to understand how these challenges manifest themselves in order to ultimately acquire a set of M67 rotation periods for which we have confidence.

We devised a series of sinusoidal injection tests with real Campaign 5 data in order to determine the best-case scenario threshold limits for K2 M67 data.  We used a subset of the SAP light curves processed via the \emph{Kepler} pipeline that fell on the spacecraft's module 6 CCD (which encompassed M67), M67 members contained in the superstamp, and a small set of hot A and F stars scattered throughout the Campaign 5 field of view as the three test samples for the injection tests.  Into each raw light curve of each sample, we injected sinusoids with six different amplitudes ranging from 0.05\% to 3.0\% and seven periods ranging from 5 to 35\,d.  We processed the injected light curves using two different methods based on existing literature, within this paper known as the `Oxford' and `CfA' pipelines.  We then defined a detection threshold and ran a Lomb-Scargle periodogram on the injected light curves and their non-injected counterparts.  We normalized the periodograms of both the injected and non-injected light curves by dividing out the median periodograms of the latter.  Those non-injected light curves whose normalized periodograms met the detection threshold were marked as `variable,' and we removed the corresponding stars from the injected samples.  Finally, we analyzed the results of the period search on the remaining sets of injected light curves in terms of completeness and reliability.  Completeness describes how sensitive our methods are in recovering the injected signals, while reliability quantifies how trustworthy a detected period is within a measured amplitude range.

The results of the injection tests shed light on the nature of the K2 M67 data.  The hot star samples from the Oxford and CfA pipelines highlighted the presence of non-astrophysical trends in the data with power at $\sim$25\,d and longer, a problematic feature --- especially given the periods we expect in M67 --- which necessitated the periodogram normalization in the period search.  In general, for the hot star, SAP, and SS samples from both pipelines, completeness diminished as the period increased and amplitude decreased.  The hot star completeness was higher than the SAP and SS samples due to the brightness of the stars in the sample and the lack of rotational modulation beyond about 10\,d for this particular dataset.  However, for the SAP and SS samples from both pipelines, the completeness falls around or below about 50\% at injected periods of 20\,d and amplitudes of 0.50\%.  Crucially, at the solar case --- where the injected period is 25\,d and the amplitude 0.10\% --- the maximum completeness is only around 15\% for the SAP and SS datasets.  Despite the generally low sensitivity, the reliability is typically very high, consistently at values greater than 90\%, except in the instance of very long periods ($\sim$35\,d) and a couple cases at detected periods of around 15\,d with moderately low measured amplitudes (0.075 -- 0.25\%).  This means that while it is very difficult to detect the kind of periods we expect to see in M67 given its age, if we do get a detection, we can generally trust the result.  

The CfA completeness is generally greater in both the SAP and SS samples.  Except at long periods and the $\sim$15\,d range, the reliability for the SAP samples from both pipelines are comparable, but the Oxford reliability is slightly higher for the SS, despite amplitude suppression from the PCA.  There is a slight trade-off between the two pipelines, and while the CfA performs marginally better overall, both are viable options.

It is important to understand how this study alters our understanding of both future and previously published studies regarding rotation periods in M67 from K2 data, namely \citet{2016ApJ...823...16B} and \citet{2016MNRAS.463.3513G}.  Overall, we urge caution when using the periods from \citet{2016ApJ...823...16B} and \citet{2016MNRAS.463.3513G} because we cannot reproduce the same results here without lowering the detection threshold and sacrificing reliability.  However, we can use the injection tests as a basis for the re-evaluation of M67 periods.  With completeness, we can quantify just how difficult is to find even the best-case, long-period rotation signals in M67 data, or how many detections we may miss, while reliability gives us confidence in our measurement.  To maintain some flexibility while still trusting the result, we decide on a reliability threshold of 80\% for a given measured period and amplitude range.

While this study has given us important insights into the K2 M67 data, we have to remember that the sinusoidal injections offer a best-case scenario.  The rotational modulation in most stars does not usually take the form of a simple sinusoid.  We ultimately want to do another, smaller round of injection tests using more realistic signals, such as those generated from the star spot models of \citet{2015MNRAS.450.3211A}.  This necessarily means that the Lomb-Scargle periodogram may not end up being the optimum rotation detection method.  As a result, we will use the star spot model tests to compare the Lomb-Scargle with the autocorrelation function and a Gaussian process.  Following this, we hope to present our own rotation periods for M67 from Campaign 5 data, informed by both the sinusoidal and star spot model injection tests, along with a comparison to previously published results and existing theory.  While we have shown that it is difficult to find low-amplitude, 25\,d periods in the K2 M67 data from Campaign 5, we should still be able to infer important information regarding the true rotation periods in M67 in a manner similar to the planet occurrence studies conducted by \citet{0067-0049-201-2-15} and \citet{2013PNAS..11019273P}.  In addition, Campaign 16, which has just been released at the time of this writing, will greatly enhance our confidence in any rotation periods measured from Campaign 5 data, as the extra observation time will allow us to improve completeness without sacrificing reliability.

We conducted this study specifically in the context of K2 Campaign 5 M67 light curves, but the caution we have urged based on our conclusions can reasonably be extended to rotation studies with other K2 campaigns and the upcoming, first data release from TESS.  We have shown that long-term, systematic trends still exist in K2 data which could skew low-amplitude period detections of about 25\,d and longer, and these trends are likely to be present in other campaigns, especially in the crowded fields of clusters.  Finally, while TESS will survey nearly the entire sky, the most coverage will be along the Ecliptic poles as opposed to the Ecliptic itself, where observations will only last about 27\,d \citep{2014SPIE.9143E..20R}.  While this will complement K2 data, the short observation time and large pixel sizes (15\,$\mu$m x 15\,$\mu$m; \citealt{2014SPIE.9143E..20R}) will likely cause the data to suffer from similar problems that we have seen in this study.

\acknowledgments
R.E. would like to thank the Rhodes Trust and the U.S. Air Force for helping finance this work.  The views expressed in this article are those of the authors and do not necessarily reflect the official policy or position of the Air Force, the Department of Defense or the U.S. Government.  Financial support for this work also includes the Science and Technology Facilities Council (STFC) consolidated grants ST/H002456/1 and ST/N000919/1.  This work was performed in part under contract with the California Institute of Technology/Jet Propulsion Laboratory funded by NASA through the Sagan Fellowship Program executed by the NASA Exoplanet Science Insitute. Some/all of the data presented in this paper were obtained from the Mikulski Archive for Space Telescopes (MAST). STScI is operated by the Association of Universities for Research in Astronomy, Inc., under NASA contract NAS5-26555.  

\appendix
\section{Completeness and Reliability Tables} \label{sec:result_tables}
The complete completeness and reliability results for the hot star, SAP, and SS samples from the Oxford and CfA pipelines are given here.  Tables~\ref{tab:hotcomplete} to \ref{tab:ss_av_complete} show the completeness results, while Tables~\ref{tab:hotreliability1} to \ref{tab:ss_av_reliability1} show the reliability results, both rounded to the nearest percent.  In the reliability tables, the number of detections for each bin is given in brackets below the reliability statistic.  Uncertainties are provided.
\begin{deluxetable*}{lccccccc}
\tablecaption{Oxford Hot Star Completeness (\%) \label{tab:hotcomplete}}
\tablecolumns{8}
\tablenum{1}
\tablewidth{0pt}
\tablehead{
\colhead{$a_{\rm inj} \setminus P_{\rm inj}$} &
\colhead{5d} &
\colhead{10d} & \colhead{15d} & \colhead{20d} &
\colhead{25d} & \colhead{30d} &
\colhead{35d}
}
\startdata
3.00\% & 98 & 95 & 95 & 94 & 94 & 92 & 88\\
1.00\% & 95 & 94 & 88 & 88 & 83 & 73 & 59\\
0.50\% & 92 & 89 & 80 & 70 & 65 & 52 & 35\\
0.30\% & 86 & 80 & 71 & 50 & 48 & 26 & 23\\
0.10\% & 65 & 47 & 27 & 17 & 17 & 11 & 3\\
0.05\% & 41 & 24 & 17 & 12 & 9 & 3 & 2\\
\enddata
\tablecomments{Total number of injections per period-amplitude bin: 66; $\sigma=\pm12.3\%$}
\end{deluxetable*}

\begin{deluxetable*}{lccccccc}
\tablecaption{CfA Hot Star Completeness (\%) \label{tab:hot_cfa_complete}}
\tablecolumns{8}
\tablenum{2}
\tablewidth{0pt}
\tablehead{
\colhead{$a_{\rm inj} \setminus P_{\rm inj}$} &
\colhead{5d} &
\colhead{10d} & \colhead{15d} & \colhead{20d} &
\colhead{25d} & \colhead{30d} &
\colhead{35d}
}
\startdata
3.00\% & 100 & 99 & 94 & 91 & 87 & 83 & 80\\
1.00\% & 97 & 93 & 91 & 84 & 81 & 77 & 76\\
0.50\% & 93 & 90 & 86 & 83 & 77 & 73 & 60\\
0.30\% & 91 & 87 & 84 & 80 & 69 & 49 & 34\\
0.10\% & 83 & 66 & 56 & 47 & 44 & 34 & 20\\
0.05\% & 66 & 50 & 36 & 16 & 16 & 13 & 6\\
\enddata
\tablecomments{Total number of injections per period-amplitude bin: 70; $\sigma=\pm12.0\%$}
\end{deluxetable*}

\begin{deluxetable*}{lccccccc}
\tablecaption{Oxford SAP Completeness (\%) \label{tab:sapcomplete}}
\tablecolumns{8}
\tablenum{3}
\tablewidth{0pt}
\tablehead{
\colhead{$a_{\rm inj} \setminus P_{\rm inj}$} &
\colhead{5d} &
\colhead{10d} & \colhead{15d} & \colhead{20d} &
\colhead{25d} & \colhead{30d} &
\colhead{35d}
}
\startdata
3.00\% & 99 & 97 & 92 & 86 & 85 & 79 & 65\\
1.00\% & 93 & 81 & 75 & 65 & 65 & 56 & 28\\
0.50\% & 82 & 71 & 58 & 43 & 43 & 30 & 8\\
0.30\% & 73 & 59 & 40 & 24 & 21 & 12 & 3\\
0.10\% & 44 & 21 & 5 & 3 & 2 & 1 & 1\\
0.05\% & 21 & 4 & 1 & 1 & 0 & 0 & 0\\
\enddata
\tablecomments{Total number of injections per period-amplitude bin: 1639; $\sigma=\pm2.5\%$}
\end{deluxetable*}

\begin{deluxetable*}{lccccccc}
\tablecaption{CfA SAP Completeness (\%) \label{tab:sap_cfa_complete}}
\tablecolumns{8}
\tablenum{4}
\tablewidth{0pt}
\tablehead{
\colhead{$a_{\rm inj} \setminus P_{\rm inj}$} &
\colhead{5d} &
\colhead{10d} & \colhead{15d} & \colhead{20d} &
\colhead{25d} & \colhead{30d} &
\colhead{35d}
}
\startdata
3.00\% & 97 & 92 & 89 & 82 & 74 & 72 & 68\\
1.00\% & 91 & 81 & 73 & 66 & 61 & 53 & 39\\
0.50\% & 83 & 69 & 63 & 53 & 47 & 33 & 23\\
0.30\% & 74 & 58 & 52 & 41 & 37 & 30 & 18\\
0.10\% & 54 & 33 & 23 & 18 & 14 & 11 & 7\\
0.05\% & 35 & 19 & 9 & 5 & 3 & 2 & 2\\
\enddata
\tablecomments{Total number of injections per period-amplitude bin: 1587; $\sigma=\pm2.5\%$}
\end{deluxetable*}

\begin{deluxetable*}{lccccccc}
\tablecaption{Oxford Superstamp Completeness (\%) \label{tab:sscomplete}}
\tablecolumns{8}
\tablenum{5}
\tablewidth{0pt}
\tablehead{
\colhead{$a_{\rm inj} \setminus P_{\rm inj}$} &
\colhead{5d} &
\colhead{10d} & \colhead{15d} & \colhead{20d} &
\colhead{25d} & \colhead{30d} &
\colhead{35d}
}
\startdata
3.00\% & 100 & 98 & 93 & 88 & 89 & 77 & 74\\
1.00\% & 91 & 81 & 74 & 65 & 66 & 50 & 46\\
0.50\% & 78 & 66 & 54 & 46 & 46 & 37 & 34\\
0.30\% & 68 & 50 & 41 & 38 & 37 & 27 & 24\\
0.10\% & 39 & 30 & 23 & 15 & 13 & 2 & 2\\
0.05\% & 28 & 14 & 8 & 3 & 2 & 0 & 0\\
\enddata
\tablecomments{Total number of injections per period-amplitude bin: 843; $\sigma=\pm3.4\%$}
\end{deluxetable*}

\begin{deluxetable*}{lccccccc}
\tablecaption{CfA Superstamp Completeness (\%) \label{tab:ss_av_complete}}
\tablecolumns{8}
\tablenum{6}
\tablewidth{0pt}
\tablehead{
\colhead{$a_{\rm inj} \setminus P_{\rm inj}$} &
\colhead{5d} &
\colhead{10d} & \colhead{15d} & \colhead{20d} &
\colhead{25d} & \colhead{30d} &
\colhead{35d}
}
\startdata
3.00\% & 93 & 86 & 83 & 74 & 63 & 56 & 51\\
1.00\% & 82 & 70 & 61 & 52 & 45 & 36 & 27\\
0.50\% & 69 & 57 & 50 & 40 & 34 & 27 & 23\\
0.30\% & 58 & 47 & 41 & 34 & 31 & 26 & 16\\
0.10\% & 42 & 31 & 21 & 15 & 15 & 11 & 7\\
0.05\% & 29 & 15 & 10 & 5 & 4 & 4 & 2\\
\enddata
\tablecomments{Total number of injections per period-amplitude bin: 844; $\sigma=\pm3.4\%$}
\end{deluxetable*}

\begin{deluxetable*}{lrrrrrrr}
\tablecaption{Oxford Hot Star Reliability (\%) \label{tab:hotreliability1}}
\tablecolumns{8}
\tablenum{7}
\tablewidth{0pt}
\tablehead{
\colhead{$2\alpha_{peak}$/$p_{peak}$} &
\colhead{5$\pm$2.5d} &
\colhead{10$\pm$2.5d} & \colhead{15$\pm$2.5d} & \colhead{20$\pm$2.5d} &
\colhead{25$\pm$2.5d} & \colhead{30$\pm$2.5d} &
\colhead{32.5d+}
}
\startdata
$\geq$2.00\% & 100 $\pm$12.4 & 98 $\pm$12.5 & 100 $\pm$12.6 & 98 $\pm$12.6 & 100 $\pm$12.7 & 100 $\pm$12.8 & 100 $\pm$13.1 \\
 & [65] & [64] & [63] & [63] & [62] & [61] & [58] \\
0.75-2.00\% & 100 $\pm$12.6 & 100 $\pm$12.7 & 98 $\pm$13.0 & 100 $\pm$13.1 & 98 $\pm$13.4 & 100 $\pm$14.1 & 100 $\pm$16.4 \\
 & [63] & [62] & [59] & [58] & [56] & [50] & [37] \\
0.45-0.75\% & 100 $\pm$13.0 & 100 $\pm$13.5 & 100 $\pm$13.9 & 100 $\pm$16.0 & 100 $\pm$16.7 & 100 $\pm$17.1 & 94 $\pm$24.3 \\
 & [59] & [55] & [52] & [39] & [36] & [34] & [17] \\
0.25-0.45\% & 98 $\pm$12.9 & 100 $\pm$13.5 & 100 $\pm$14.4 & 97 $\pm$16.0 & 100 $\pm$16.0 & 100 $\pm$22.9 & 62 $\pm$17.7 \\
 & [60] & [55] & [48] & [39] & [39] & [19] & [32] \\
0.075-0.25\% & 93 $\pm$14.7 & 100 $\pm$17.4 & 100 $\pm$23.6 & 100 $\pm$26.7 & 100 $\pm$30.2 & 100 $\pm$37.8 & 100 $\pm$70.7\\
 & [46] & [33] & [18] & [14] & [11] & [7] & [2] \\
$<$0.075\% & 93 $\pm$18.6 & 100 $\pm$25.0 & 100 $\pm$30.2 & 100 $\pm$37.8 & 100 $\pm$37.8 & 100 $\pm$70.7 & -- \\
 & [29] & [16] & [11] & [7] & [7] & [2] & [0] \\
\enddata
\end{deluxetable*}

\begin{deluxetable*}{lrrrrrrr}
\tablecaption{CfA Hot Star Reliability (\%) \label{tab:hot_av_reliability1}}
\tablecolumns{8}
\tablenum{8}
\tablewidth{0pt}
\tablehead{
\colhead{$2\alpha_{peak}$/$p_{peak}$} &
\colhead{5$\pm$2.5d} &
\colhead{10$\pm$2.5d} & \colhead{15$\pm$2.5d} & \colhead{20$\pm$2.5d} &
\colhead{25$\pm$2.5d} & \colhead{30$\pm$2.5d} &
\colhead{32.5d+}
}
\startdata
$\geq$2.00\% & 100 $\pm$12.0 & 100 $\pm$12.0 & 100 $\pm$12.3 & 100 $\pm$12.7 & 100 $\pm$13. & 100 $\pm$13.4 & 100 $\pm$13.4 \\
 & [70] & [69] & [66] & [62] & [59] & [56] & [56] \\
0.75-2.00\% & 100 $\pm$12.1 & 98 $\pm$12.3 & 100 $\pm$12.5 & 100 $\pm$13.0 & 100 $\pm$13.6 & 100 $\pm$13.2 & 100 $\pm$13.7 \\
 & [68] & [66] & [64] & [59] & [54] & [57] & [53] \\
0.45-0.75\% & 100 $\pm$12.5 & 100 $\pm$12.7 & 100 $\pm$13.0 & 100 $\pm$13.2 & 100 $\pm$13.9 & 100 $\pm$15.1 & 100 $\pm$17.1 \\
 & [64] & [62] & [59] & [57] & [52] & [44] & [34] \\
0.25-0.45\% & 100 $\pm$12.4 & 100 $\pm$12.9 & 100 $\pm$13.0 & 100 $\pm$13.1 & 100 $\pm$14.1 & 98 $\pm$15.6 & 100 $\pm$18.0 \\
 & [65] & [60] & [59] & [58] & [50] & [41] & [31] \\
0.075-0.25\% & 95 $\pm$12.8 & 100 $\pm$14.3 & 98 $\pm$15.2 & 100 $\pm$17.4 & 100 $\pm$18.9 & 100 $\pm$18.9 & 87 $\pm$25.8 \\
 & [61] & [49] & [43] & [33] & [28] & [28] & [15] \\
$<$0.075\% & 84 $\pm$13.5 & 97 $\pm$16.9 & 96 $\pm$20.4 & 93 $\pm$25.8 & 100 $\pm$27.7 & 100 $\pm$28.9 & 100 $\pm$57.7 \\
 & [55] & [35] & [24] & [15] & [13] & [12] & [3] \\
\enddata
\end{deluxetable*}

\begin{deluxetable*}{lrrrrrrr}
\tablecaption{Oxford SAP Reliability (\%) \label{tab:sapreliability1}}
\tablecolumns{8}
\tablenum{9}
\tablewidth{0pt}
\tablehead{
\colhead{$2\alpha_{peak}$/$p_{peak}$} &
\colhead{5$\pm$2.5d} &
\colhead{10$\pm$2.5d} & \colhead{15$\pm$2.5d} & \colhead{20$\pm$2.5d} &
\colhead{25$\pm$2.5d} & \colhead{30$\pm$2.5d} &
\colhead{32.5d+}
}
\startdata
$\geq$2.00\% & 100 $\pm$2.5 & 99 $\pm$2.5 & 99 $\pm$2.6 & 100 $\pm$2.7 & 99 $\pm$2.7 & 99 $\pm$2.6 & 96 $\pm$3.2 \\
 & [1620] & [1598] & [1525] & [1414] & [1405] & [1440] & [996] \\
0.75-2.00\% & 100 $\pm$2.6 & 98 $\pm$2.7 & 99 $\pm$2.8 & 100 $\pm$3.0 & 98 $\pm$3.0 & 99 $\pm$3.1 & 88 $\pm$4.7 \\
 & [1523] & [1364] & [1250] & [1076] & [1095] & [1025] & [456] \\
0.45-0.75\% & 100 $\pm$2.8 & 98 $\pm$3.1 & 98 $\pm$3.5 & 100 $\pm$4.3 & 98 $\pm$4.0 & 98 $\pm$4.5 & 46 $\pm$6.2 \\
 & [1267] & [1065] & [812] & [534] & [623] & [495] & [261] \\
0.25-0.45\% & 98 $\pm$2.8 & 98 $\pm$3.1 & 98 $\pm$3.6 & 100 $\pm$4.4 & 100 $\pm$5.1 & 99 $\pm$6.8 & 33 $\pm$7.4 \\
 & [1278] & [1074] & [768] & [519] & [391] & [214] & [184] \\
0.075-0.25\% & 95 $\pm$3.5 & 97 $\pm$5.1 & 85 $\pm$7.8 & 99 $\pm$10.3 & 98 $\pm$14.4 & 95 $\pm$22.9 & 20 $\pm$14.3 \\
 & [832] & [386] & [165] & [95] & [48] & [19] & [49] \\
$<$0.075\% & 83 $\pm$5.3 & 97 $\pm$12.4 & 100 $\pm$22.4 & 100 $\pm$25.0 & 100 $\pm$44.7 & 100 $\pm$100.0 & 100 $\pm$100.0 \\
 & [358] & [65] & [20] & [16] & [5] & [1] & [1] \\
\enddata
\end{deluxetable*}

\begin{deluxetable*}{lrrrrrrr}
\tablecaption{CfA SAP Reliability (\%) \label{tab:sap_av_reliability1}}
\tablecolumns{8}
\tablenum{10}
\tablewidth{0pt}
\tablehead{
\colhead{$2\alpha_{peak}$/$p_{peak}$} &
\colhead{5$\pm$2.5d} &
\colhead{10$\pm$2.5d} & \colhead{15$\pm$2.5d} & \colhead{20$\pm$2.5d} &
\colhead{25$\pm$2.5d} & \colhead{30$\pm$2.5d} &
\colhead{32.5d+}
}
\startdata
$\geq$2.00\% & 98 $\pm$2.5 & 95 $\pm$2.5 & 99 $\pm$2.6 & 100 $\pm$2.8 & 97 $\pm$2.9 & 98 $\pm$2.9 & 100 $\pm$3.1 \\
 & [1567] & [1539] & [1426] & [1293] & [1194] & [1172] & [1070] \\
0.75-2.00\% & 100 $\pm$2.6 & 98 $\pm$2.8 & 96 $\pm$2.8 & 99 $\pm$3.0 & 99 $\pm$3.2 & 96 $\pm$3.3 & 92 $\pm$3.9 \\
 & [1470] & [1316] & [1245] & [1078] & [954] & [903] & [667] \\
0.45-0.75\% & 98 $\pm$2.8 & 97 $\pm$3.1 & 93 $\pm$3.1 & 95 $\pm$3.5 & 96 $\pm$3.8 & 92 $\pm$4.2 & 90 $\pm$5.5 \\
 & [1313] & [1057] & [1011] & [797] & [677] & [567] & [333] \\
0.25-0.45\% & 98 $\pm$2.9 & 93 $\pm$3.1 & 95 $\pm$3.3 & 98 $\pm$3.7 & 96 $\pm$4.0 & 93 $\pm$4.0 & 82 $\pm$5.2 \\
 & [1213] & [1050] & [896] & [721] & [640] & [618] & [365] \\
0.075-0.25\% & 96 $\pm$3.3 & 98 $\pm$4.2 & 91 $\pm$4.8 & 98 $\pm$5.5 & 95 $\pm$6.8 & 94 $\pm$6.5 & 75 $\pm$7.8 \\
 & [915] & [575] & [426] & [329] & [217] & [237] & [165] \\
$<$0.075\% & 95 $\pm$4.2 & 95 $\pm$5.9 & 95 $\pm$8.6 & 100 $\pm$13.4 & 88 $\pm$15.2 & 100 $\pm$16.7 & 93 $\pm$26.7 \\
 & [572] & [291] & [134] & [56] & [43] & [36] & [14] \\
\enddata
\end{deluxetable*}

\begin{deluxetable*}{lrrrrrrr}
\tablecaption{Oxford Superstamp Reliability (\%) \label{tab:ssreliability1}}
\tablecolumns{8}
\tablenum{11}
\tablewidth{0pt}
\tablehead{
\colhead{$2\alpha_{peak}$/$p_{peak}$} &
\colhead{5$\pm$2.5d} &
\colhead{10$\pm$2.5d} & \colhead{15$\pm$2.5d} & \colhead{20$\pm$2.5d} &
\colhead{25$\pm$2.5d} & \colhead{30$\pm$2.5d} &
\colhead{32.5d+}
}
\startdata
$\geq$2.00\% & -- & 100 $\pm$100.0 & 75 $\pm$25.0 & 35 $\pm$15.8 & 100 $\pm$27.7 & 86 $\pm$37.8 & 36 $\pm$13.4 \\
 & [0] & [1] & [16] & [40] & [13] & [7] & [56] \\
0.75-2.00\% & 95 $\pm$3.4 & 100 $\pm$3.5 & 99 $\pm$3.5 & 99 $\pm$3.6 & 100 $\pm$3.6 & 99 $\pm$3.9 & 99 $\pm$4.0 \\
 & [886] & [826] & [797] & [752] & [760] & [657] & [620] \\
0.45-0.75\% & 100 $\pm$3.7 & 100 $\pm$3.9 & 98 $\pm$4.1 & 96 $\pm$4.5 & 95 $\pm$4.5 & 99 $\pm$6.8 & 92 $\pm$7.2 \\
 & [739] & [652] & [584] & [497] & [501] & [215] & [191]\\
0.25-0.45\% & 92 $\pm$4.9 & 100 $\pm$5.7 & 99 $\pm$6.0 & 99 $\pm$6.6 & 98 $\pm$6.4 & 99 $\pm$6.0 & 96 $\pm$6.2 \\
 & [421] & [311] & [279] & [227] & [246] & [274] & [262] \\
0.075-0.25\% & 98 $\pm$3.3 & 98 $\pm$3.7 & 97 $\pm$4.1 & 99 $\pm$4.2 & 99 $\pm$4.2 & 99 $\pm$4.6 & 96 $\pm$4.8 \\
 & [898] & [713] & [601] & [558] & [563] & [478] & [442] \\
$<$0.075\% & 99 $\pm$4.2 & 96 $\pm$5.1 & 96 $\pm$6.3 & 94 $\pm$8.3 & 97 $\pm$9.4 & 84 $\pm$20.0 & 70 $\pm$22.4 \\
 & [557] & [388] & [253] & [145] & [114] & [25] & [20] \\
\enddata
\end{deluxetable*}

\begin{deluxetable*}{lrrrrrrr}
\tablecaption{CfA Superstamp Reliability (\%) \label{tab:ss_av_reliability1}}
\tablecolumns{8}
\tablenum{12}
\tablewidth{0pt}
\tablehead{
\colhead{$2\alpha_{peak}$/$p_{peak}$} &
\colhead{5$\pm$2.5d} &
\colhead{10$\pm$2.5d} & \colhead{15$\pm$2.5d} & \colhead{20$\pm$2.5d} &
\colhead{25$\pm$2.5d} & \colhead{30$\pm$2.5d} &
\colhead{32.5d+}
}
\startdata
$\geq$2.00\% & 97 $\pm$3.5 & 95 $\pm$3.6 & 99 $\pm$3.8 & 100 $\pm$4.0 & 93 $\pm$4.3 & 96 $\pm$4.4 & 97 $\pm$4.9 \\
 & [810] & [768] & [703] & [620] & [545] & [513] & [411] \\
0.75-2.00\% & 94 $\pm$3.7 & 96 $\pm$4.0 & 87 $\pm$4.0 & 98 $\pm$4.8 & 94 $\pm$5.0 & 91 $\pm$5.0 & 80 $\pm$6.4 \\
 & [740] & [619] & [613] & [428] & [394] & [405] & [241] \\
0.45-0.75\% & 97 $\pm$4.2 & 95 $\pm$4.6 & 91 $\pm$4.9 & 91 $\pm$5.5 & 82 $\pm$6.3 & 88 $\pm$5.8 & 72 $\pm$8.8 \\
 & [568] & [477] & [418] & [329] & [256] & [302] & [128]\\
0.25-0.45\% & 87 $\pm$4.1 & 92 $\pm$4.8 & 89 $\pm$4.9 & 96 $\pm$5.7 & 84 $\pm$6.0 & 87 $\pm$5.0 & 75 $\pm$7.4 \\
 & [587] & [434] & [418] & [312] & [278] & [396] & [183] \\
0.075-0.25\% & 93 $\pm$5.0 & 79 $\pm$5.3 & 82 $\pm$6.6 & 94 $\pm$7.1 & 98 $\pm$8.5 & 93 $\pm$8.0 & 67 $\pm$10.6 \\
 & [400] & [356] & [232] & [199] & [139] & [158] & [89] \\
$<$0.075\% & 97 $\pm$6.4 & 84 $\pm$8.4 & 97 $\pm$11.9 & 96 $\pm$14.9 & 91 $\pm$20.9 & 95 $\pm$21.8 & 80 $\pm$25.8 \\
 & [241] & [141] & [71] & [45] & [23] & [21] & [15] \\
\enddata
\end{deluxetable*}

\section{Additional Experiments} \label{sec:pubcompare}
Here we present the results from additional experiments we conducted to further understand the benefits of normalization of the Lomb-Scargle periodogram light curves and to compare our results for the non-injected versions of the light curves with those published in \citet{2016ApJ...823...16B} and \citet{2016MNRAS.463.3513G}.  In \ref{subsec:nonorm}, we show in Figure~\ref{fig:nonorm1} the completeness and reliability values for the Oxford hot star and SAP datasets without normalization of the Lomb-Scargle periodograms compared to the results from the normalized versions.  In \ref{subsec:detthresh}, we show the measured period from the Oxford and CfA pipelines for the EPICs listed in \citet{2016ApJ...823...16B} and the associated values of $C$ needed to count them as detections.

\subsection{Completeness and Reliability without Normalization} \label{subsec:nonorm}

\begin{figure*}
\includegraphics[width=1\linewidth]{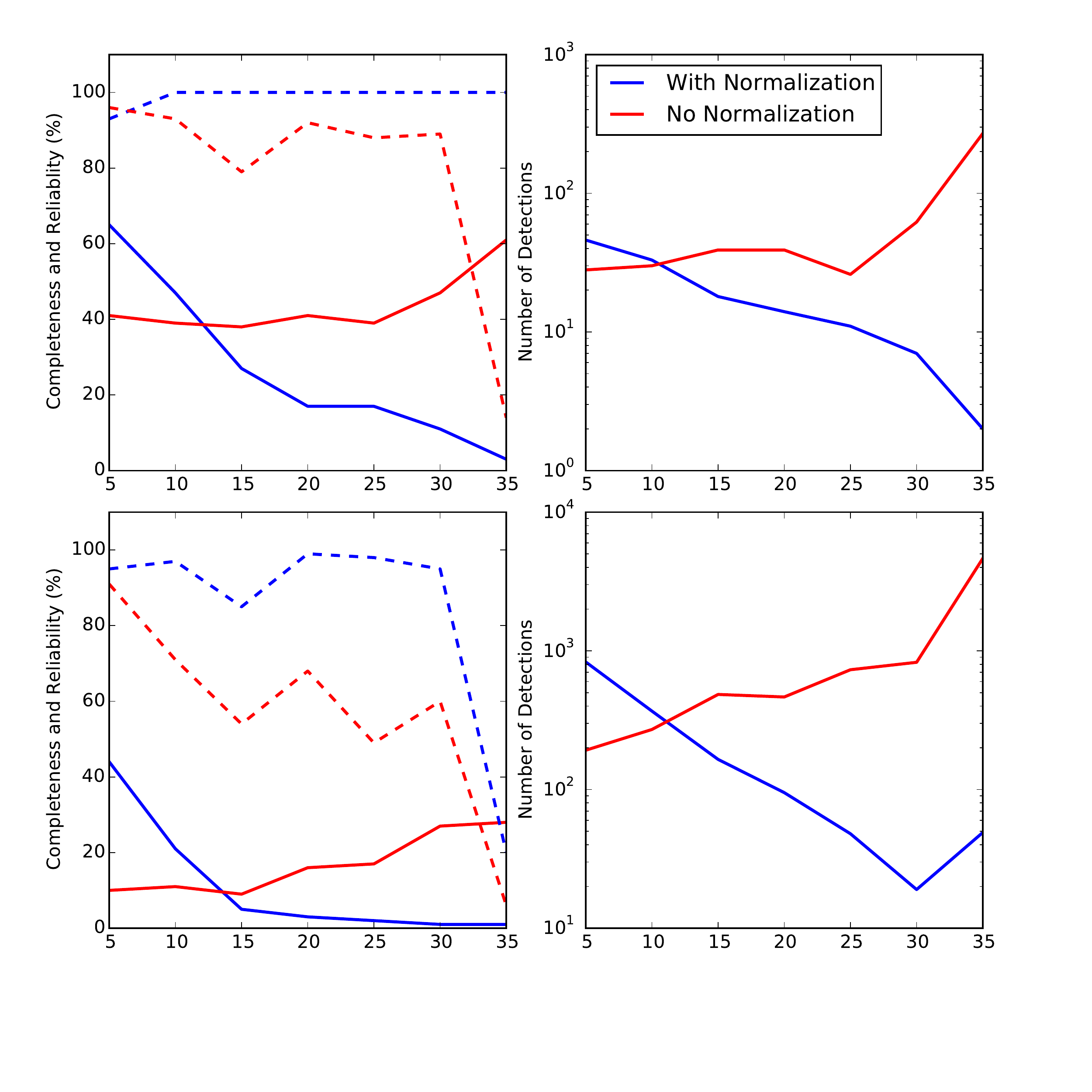} \caption{Comparison of completeness and reliability results for the Oxford hot star (top row) and SAP (bottom row) datasets with (blue) and without (red) normalization of the Lomb-Scargle periodogram for the solar amplitude (0.1\% for completeness, 0.07--0.25\% range for reliability).  The left panels show the completeness (solid lines) and reliability (dashed lines) results, while the right panels show the number of detected periods with measured amplitudes of 0.07--0.25\% (i.e. similar to the numbers in brackets in Tables~\ref{tab:hotreliability1} through \ref{tab:ss_av_reliability1}).  \label{fig:nonorm1}}
\end{figure*}

To see the effect of using the regular Lomb-Scargle periodogram (i.e. without normalization) on the completeness and reliability for the K2 Campaign 5 M67 light curves, we computed the periods for the Oxford hot star and SAP injected datasets without using the normalized power.  For a detection threshold, we followed the example of \citet{2013A&A...557L..10N} and set the threshold at four times the root mean square (RMS) of the zero-mean time series.  We removed the same stars marked as variables as in the main set of injection tests.  Figure~\ref{fig:nonorm1} shows the completeness and reliability results for the solar amplitude (0.1\% for completeness, 0.07--0.25\% for reliability), along with the number of detections with measured amplitudes ranging from 0.07 to 0.25\% (i.e. similar to the numbers in brackets in Tables~\ref{tab:hotreliability1} through \ref{tab:ss_av_reliability1}).  The blue lines show the results with normalization, while the red lines show the results without normalization.  Due to the lower detection threshold criteria and the lingering presence of the long-term trends seen in the median periodograms, the number of detections for the un-normalized samples increases significantly compared to the normalized hot star and SAP samples.  This allows for an increase in completeness, including at the solar case, which improved from 17\% with the normalized periodograms to 39\% for the regular Lomb-Scargle with the hot stars, and from 2\% to 17\% for the SAP sample.  In addition, completeness begins to increase with increasing period within an injected amplitude starting around 25\,d, countering the general trends of the normalized sample.  However, as expected, the reliability at periods of 25\,d and longer decreases drastically.  There is also a disproportionate number of detections at these longer periods, showing just how much of an effect the lingering long-term systematic features have on periodic measurements.

\subsection{Detection Thresholds and Periods for Barnes et al. (2016) Stars} \label{subsec:detthresh}

\begin{deluxetable*}{lcccccccc}
\tablecaption{Evaluation of \citet{2016ApJ...823...16B} \label{tab:barnes_compare2}}
\tablecolumns{8}
\tablenum{16}
\tablewidth{0pt}
\tablehead{
\colhead{EPIC} &
\colhead{Barnes Per (d)} &
\colhead{Ox Per (d)} & \colhead{Ox $C$} & \colhead{Ox Rel (\%)} & \colhead{CfA Per (d)} & \colhead{CfA $C$} & \colhead{CfA Rel (\%)}
}
\startdata
211388204 & 31.8 & 2.2 & 1 & 6 $\pm$3.2 & 2.1 & 1 & 26 $\pm$8.0\\
211394185 & 30.4 & 15.4 & 4 & 98 $\pm$3.6 & 15.1 & 4 & 95 $\pm$3.3\\
211395620 & 30.7 & 28.5 & 2 & 98 $\pm$4.5 & 3.5 & 2 & 38 $\pm$3.3\\
211397319 & 25.1 & 7.9 & 2 & 96 $\pm$12.0 & 47.6 & 1 & 54 $\pm$16.0\\
211397512 & 34.5 & 16.4 & 4 & 98 $\pm$3.6 & 38.4 & 3 & 85 $\pm$5.7\\
211398025 & 28.8 & 22.2 & 2 & 100 $\pm$28.9 & 10.4 & 2 & 78 $\pm$4.5\\
211398541 & 30.3 & 5.3 & -- & -- & 16.9 & 2 & 77 $\pm$5.0\\
211399458 & 30.2 & 7.9 & 2 & 86 $\pm$4.5 & 4.32 & 1 & 17 $\pm$5.2\\
211399819 & 28.4 & 13.9 & 2 & 36 $\pm$2.4 & 2.1 & 1 & 17 $\pm$5.2\\
211400500 & 26.9 & 13.5 & 3 & 82 $\pm$6.6 & 2.0 & 2 & 38 $\pm$3.3\\
211406596 & 26.9 & 31.2 & 2 & 85 $\pm$6.9 & 2.4 & 1 & 26 $\pm$8.0\\
211410757 & 18.9 & 18.8 & -- & -- & 9.4 & 3 & 93 $\pm$5.6\\
211411477 & 31.2 & 5.7 & 1 & 6 $\pm$3.2 & 5.7 & 1 & 26 $\pm$8.0\\
211411621 & 30.5 & 15.8 & 4 & 85 $\pm$7.8 & 16.1 & 4 & 91 $\pm$4.8\\
211413212 & 24.4 & 5.3 & 1 & 6 $\pm$3.2 & 7.5 & 1 & 61 $\pm$14.7\\
211413961 & 31.4 & 2.9 & 1 & 6 $\pm$3.2 & 15.4 & 3 & 88 $\pm$4.5\\
211414799 & 18.1 & 4.6 & 3 & 100 $\pm$20.9 & 9.0 & 4 & 98 $\pm$4.2\\
211423010 & 24.9 & 20.8 & 2 & 92 $\pm$7.8 & 22.2 & 4 & 98 $\pm$5.5\\
211428580 & 26.9 & 5.1 & 1 & 6 $\pm$3.2 & 2.4 & 1 & 26 $\pm$8.0\\
211430274 & 31.1 & 23.2 & 2 & 89 $\pm$11.2 & 27.0 & 2 & 93 $\pm$6.8\\
\enddata
\tablecomments{The columns from left to right are EPIC, the associated period reported from \citet{2016ApJ...823...16B}, the period from the Oxford pipeline, the Oxford threshold factor $C$ that results in a detection, the corresponding value for reliability, the period from the CfA pipeline, the CfA detection value for $C$, and the CfA reliability.}
\end{deluxetable*}

Table~\ref{tab:barnes_compare2} shows the minimum detection threshold factors for the periods found by the normalized Lomb-Scargle periodogram to count as detections using the light curves from the Oxford and CfA pipelines for the stars in \citet{2016ApJ...823...16B}.  We present the EPICs from \citet{2016ApJ...823...16B}, the published periods, the periods computed from the normalized periodograms of the corresponding non-injected Oxford and CfA pipeline light curves, the threshold factor $C$ required to count the Oxford and CfA periods as detections, and the associated reliabilities.  In most cases, the value of $C$ had to be reduced to 2 or 1 in order for the measured period from the Oxford and CfA pipelines to be counted as a detection, which means the reliability falls considerably; the average Oxford reliability is 65\%, and the average CfA reliability is 61\%.  In addition, the Oxford and CfA periods rarely match the published values (though in some cases they are harmonics) and they do not always match each other.  This continues to show how difficult it is to find accurate rotation periods in the K2 Campaign 5 M67 light curves. 

\bibliography{techniques_bib}

\end{document}